\documentclass[pdflatex,sn-vancouver,Numbered]{sn-jnl}% Vancouver Reference Style
\usepackage{graphicx}%
\usepackage{multirow}%
\usepackage{amsmath,amssymb,amsfonts}%
\usepackage{amsthm}%
\usepackage{mathrsfs}%
\usepackage[title]{appendix}%
\usepackage[dvipsnames]{xcolor}
\usepackage{textcomp}%
\usepackage{manyfoot}%
\usepackage{booktabs}%
\usepackage{algorithm}%
\usepackage{algorithmicx}%
\usepackage{algpseudocode}%
\usepackage{listings}%
\usepackage[T1]{fontenc}
\usepackage{type1cm}
\usepackage{anyfontsize}
\usepackage{lmodern}
\usepackage{mathrsfs}
\usepackage{silence}
\usepackage{microtype}
\usepackage{hyperref}
\usepackage{float}
\usepackage{tabularx}
\usepackage[normalem]{ulem}
\usepackage{caption} % For more complex captioning
\usepackage{manyfoot}%
\usepackage{subcaption}

\newcounter{extendeddatafig}
\renewcommand{\theextendeddatafig}{\textbf{Extended Data Figure \arabic{extendeddatafig}}}

\makeatletter

\newcommand*\H@extendeddatafig{\@nameuse{extdatafigureautorefname}}
\makeatother

\newcounter{extendeddatatable}
\renewcommand{\theextendeddatatable}{\textbf{Extended Data Table \arabic{extendeddatatable}}}

\makeatletter

\newcommand*\H@extendeddatatable{\@nameuse{extdatatableautorefname}}
\makeatother

\newtheoremstyle{thmstyleone}
  {3pt} % Space above
  {3pt} % Space below
  {\itshape} % Body font
  {} % Indent amount
  {\bfseries} % Theorem head font
  {.} % Punctuation after theorem head
  {.5em} % Space after theorem head
  {} % Theorem head spec (can be left empty)

\theoremstyle{thmstyleone}
\newtheoremstyle{thmstyletwo}
  {3pt} % Space above
  {3pt} % Space below
  {\normalfont} % Body font
  {} % Indent amount
  {\bfseries} % Theorem head font
  {.} % Punctuation after theorem head
  {.5em} % Space after theorem head
  {} % Theorem head spec (can be left empty)

\theoremstyle{thmstyletwo}

\newtheoremstyle{thmstylethree}
  {3pt} % Space above
  {3pt} % Space below
  {\slshape} % Body font
  {} % Indent amount
  {\bfseries\scshape} % Theorem head font
  {.} % Punctuation after theorem head
  {.5em} % Space after theorem head
  {} % Theorem head spec

\theoremstyle{thmstylethree}

\begin{document}

\title[JWST occultation reveals unforeseen complexity in Chariklo’s ring system]{JWST occultation reveals unforeseen complexity in Chariklo’s ring system}
%\title[James Webb Space Telescope’s stellar occultation reveals unexpected phenomena in Centaur Chariklo’s ring system]{James Webb Space Telescope’s stellar occultation reveals unexpected phenomena in Centaur Chariklo’s ring system}
%\title[James Webb Space Telescope finds striking differences between the two rings of the Centaur Chariklo]{James Webb Space Telescope finds striking differences between the two rings of the Centaur Chariklo}
%\title[The James Webb Space Telescope reveals different properties for the two rings of the Centaur Chariklo]{The James Webb Space Telescope reveals different properties for the two rings of the Centaur Chariklo}

\author*[1]{\fnm{Pablo} \sur{Santos-Sanz}}\email{psantos@iaa.es}
\author[2,3,4]{\fnm{Altair R.} \sur{Gomes-Júnior}}
\author[5,6,3]{\fnm{Bruno E.} \sur{Morgado}}
\author[1,7]{\fnm{Yucel} \sur{Kilic}}
\author[8,9,10]{\fnm{Csilla} \sur{Kalup}}
\author[8,9,10]{\fnm{Csaba} \sur{Kiss}}
\author[6,3]{\fnm{Chrystian L.} \sur{Pereira}}
\author[11]{\fnm{Bryan J.} \sur{Holler}}
\author[1]{\fnm{Nicolás} \sur{Morales}}
\author[1]{\fnm{José Luis} \sur{Ortiz}}
\author[12]{\fnm{Bruno} \sur{Sicardy}}
\author[1]{\fnm{Juan Luis} \sur{Rizos}} 
\author[11]{\fnm{John} \sur{Stansberry}}
\author[13]{\fnm{Richard G.} \sur{French}}
\author[14]{\fnm{Heidi B.} \sur{Hammel}}
\author[15]{\fnm{Zhong-Yi} \sur{Lin}}
\author[7,16]{\fnm{Damya} \sur{Souami}}
\author[12,17]{\fnm{Josselin} \sur{Desmars}}
\author[18]{\fnm{Stefanie N.} \sur{Milam}}
% Contributors
\author[19,7]{\fnm{Felipe} \sur{Braga-Ribas}}
\author[5,3]{\fnm{Marcelo} \sur{Assafin}}
\author[3,4]{\fnm{Gustavo} \sur{Benedetti-Rossi}}
\author[6,3]{\fnm{Julio I.B.} \sur{Camargo}}
\author[1]{\fnm{Rene} \sur{Duffard}}
\author[20]{\fnm{Flavia L.} \sur{Rommel}}
\author[20]{\fnm{Estela} \sur{Fernández-Valenzuela}}
\author[20]{\fnm{Noemí} \sur{Pinilla-Alonso}}
\author[1]{\fnm{Mónica} \sur{Vara-Lubiano}}

\affil[1]{\orgdiv{Instituto de Astrof\'{\i}sica de Andaluc\'{\i}a}, \orgname{CSIC}, \orgaddress{\street{Glorieta de la Astronom\'{\i}a s/n}, \city{Granada}, \postcode{18008}, \country{Spain}}}

\affil[2]{\orgdiv{Instituto de Física, Universidade Federal de Uberlândia (UFU)}, \orgaddress{Av. João Naves de Ávila 2121, Bairro Santa Mônica, Uberlândia, MG}, \country{Brazil}}

\affil[3]{\orgdiv{Laboratório Interinstitucional de e-Astronomia - LIneA}, \orgaddress{\street{Av. Pastor Martin Luther King Jr 126}, \postcode{20765-000}, \city{Rio de Janeiro}, \state{RJ}, \country{Brazil}}}

\affil[4]{\orgdiv{Grupo de Dinâmica Orbital e Planetologia, Faculdade de Engenharia e Ciêncas de Guaratinguetá}, \orgname{Universidade Estadual Julio de Mesquita Filho (GDOP/FEG/UNESP)}, \orgaddress{\city{Guaratinguetá}, \country{Brazil}}}

\affil[5]{\orgdiv{Federal University of Rio de Janeiro - Valongo Observatory}, \orgaddress{\city{Rio de Janeiro}, \country{Brazil}}}

\affil[6]{\orgdiv{Observatório Nacional/MCTI}, \orgaddress{\street{R. General José Cristino 77}, \postcode{20921-400}, \city{Rio de Janeiro - RJ}, \country{Brazil}}}

\affil[7]{\orgdiv{LIRA, Observatoire de Paris}, \orgname{PSL Research University, CNRS, Sorbonne Université, Univ. Paris Diderot, Sorbonne Paris Cité, 92190 Meudon}, \orgaddress{\country{France}}}

\affil[8]{\orgdiv{Konkoly Observatory, HUN-REN Research Centre for Astronomy and Earth Sciences}, \orgaddress{\street{Konkoly Thege 15-17}, \postcode{H-1121}, \city{Budapest}, \country{Hungary}}}

\affil[9]{\orgdiv{CSFK, MTA Centre of Excellence}, \orgaddress{\street{Konkoly Thege Miklós út 15-17}, \postcode{H-1121}, \city{Budapest}, \country{Hungary}}}

\affil[10]{\orgdiv{ELTE Eötvös Loránd University, Institute of Physics and Astronomy}, \orgaddress{\street{Pázmány P. st. 1/A}, \postcode{1171}, \city{Budapest}, \country{Hungary}}}

\affil[11]{\orgdiv{Space Telescope Science Institute}, \orgaddress{\city{Baltimore}, \state{Maryland}, \country{USA}}}

\affil[12]{\orgdiv{LTE, Observatoire de Paris}, \orgname{PSL Research University, Sorbonne Université, Université de Lille, LNE, CNRS, 61 Avenue de l'Observatoire, 75014 Paris}, \orgaddress{\country{France}}}

\affil[13]{\orgdiv{Department of Astronomy}, \orgname{Wellesley College}, \orgaddress{\country{USA}}}

\affil[14]{\orgdiv{AURA}, \orgaddress{\street{1331 Pennsylvania Ave NW, Suite 1475}, \city{Washington DC}, \postcode{20004}, \country{USA}}}

\affil[15]{\orgdiv{Institute of Astronomy}, \orgname{National Central University}, \orgaddress{Taoyuan City}, \country{Taiwan}}

%\affil[16]{\orgdiv{Departments of Astronomy and of Earth and Planetary Science}, \orgaddress{\street{501 Campbell Hall}, \city{Berkeley}, \state{California}, \postcode{94720}, \country{USA}}}

\affil[16]{\orgdiv{naXys, Department of Mathematics}, \orgname{University of Namur}, \orgaddress{\street{Rue de Bruxelles 61}, \postcode{B-5000}, \city{Namur}, \country{Belgium}}}

\affil[17]{\orgdiv{Institut Polytechnique des Sciences Avancées (IPSA)}, \orgaddress{\street{63 boulevard de Brandebourg}, \postcode{F-94200}, \city{Ivry-sur-Seine}, \country{France}}}

\affil[18]{\orgdiv{Goddard Space Flight Center}, \orgaddress{\city{Greenbelt}, \state{MD}, \postcode{20771}, \country{USA}}}

\affil[19]{\orgdiv{Federal University of Technology – Paraná (UTFPR)}, \orgaddress{Rua Sete de Setembro, 3165, CEP 80230-901, Curitiba, PR}, \country{Brazil}}

\affil[20]{\orgdiv{Florida Space Institute, University of Central Florida}, \orgaddress{12354 Research Parkway, Orlando, FL 32765}, \country{USA}}

%\affil[21]{\orgdiv{The Scientific and Technological Research Council of T\"{u}rkiye (T\"{U}B\.{I}TAK)}, \orgaddress{\street{Tunus Caddesi, Kavakl{\i}dere}, \postcode{06100}, \city{Ankara}, \country{T\"{u}rkiye}}}

%\author*[1]{\fnm{Pablo} \sur{Author}}\email{iauthor@gmail.com}



% \affil*[1]{\orgdiv{Department}, \orgname{Organization}, \orgaddress{\street{Street}, \city{City}, \postcode{100190}, \state{State}, \country{Country}}}

% \affil[2]{\orgdiv{Department}, \orgname{Organization}, \orgaddress{\street{Street}, \city{City}, \postcode{10587}, \state{State}, \country{Country}}}

% \affil[3]{\orgdiv{Department}, \orgname{Organization}, \orgaddress{\street{Street}, \city{City}, \postcode{610101}, \state{State}, \country{Country}}}


\abstract{
Ring systems have been discovered around several small bodies in the outer Solar System through stellar occultations. While such measurements provide key information about ring geometry and dynamical interactions, little is known about their origins, lifetimes, evolutionary pathways, or compositions. Here we report near-infrared observations with the James Webb Space Telescope (JWST) of a stellar occultation by (10199) Chariklo, a Centaur known to host a double-ring system. Our JWST measurements show that Chariklo’s inner dense ring has become significantly more opaque than in previous observations, pointing to ongoing replenishment processes or dynamical restructuring. In contrast, the outer ring exhibits a much weaker near-infrared occultation signature than seen in earlier visible-light detections. This discrepancy may reflect material loss, suggesting that the outer ring could be transient, or may arise from wavelength-dependent opacity. These scenarios, which are not mutually exclusive, point to an unprecedented level of complexity in small-body ring systems, distinct from those observed around any other minor bodies in the Solar System.

}


\maketitle

 Stellar occultations are a powerful tool in planetary astronomy, offering unique opportunities to probe the properties of distant solar system bodies and their environments with exceptional precision. When a celestial object, such as an asteroid, passes in front of a star, the brief dimming of starlight reveals information about the occulting body’s size, shape, and surroundings \cite{Sicardy2011, Sicardy2006}. This technique has also enabled the detection and characterisation of tenuous atmospheres \cite{Sicardy2016, Sicardy2021, MarquesOliveira2022}, or has placed stringent upper limits on their presence \cite{Sicardy2006}. Notably, stellar occultations have led to the discovery and study of ring systems not only around the giant planets \citep{Sicardy2024}, but more recently around small bodies, including the Centaurs Chariklo \citep{Braga-Ribas2014} and Chiron \citep{Ortiz2015, Sickafoose2020, Ortiz2023}, the dwarf planet Haumea \citep{Ortiz2017}, and the trans-Neptunian object Quaoar \citep{Morgado2023, Pereira2023}. Among these, Chariklo was the first small Solar System body discovered to host a ring system, composed of two sharply confined components, C1R and C2R, orbiting at 390 and 405~km from its centre, respectively \citep{Braga-Ribas2014, Berard2017, Morgado2021}. 
 Yet, despite extensive occultation campaigns and multiple studies on ring dynamics \citep{Sicardy2019, Sickafoose2024, Regaly2025}, the origin, composition and grain-size distribution of the rings have remained poorly constrained and have defied a clear explanation for over a decade. Although wavelength-dependent optical properties of the grains, which lead to different occultation opacities in the visible and near-infrared, may provide direct constraints on grain properties \cite{bohren2008absorption,Kalup2024}, due to the limited number of well-resolved observations and the possible structural changes in the ring, it still remains challenging.
 %Yet, despite extensive occultation campaigns and multiple studies on ring dynamics \citep{Sicardy2019, Sickafoose2024, Regaly2025}, the composition, grain-size distribution and origin of the rings have remained poorly constrained and have defied a clear explanation for over a decade. Constraining the composition of rings around small Solar System bodies remains challenging, as their proximity and faintness preclude the isolation of ring spectra using current remote spectroscopy \cite{Duffard2014}. However, wavelength-dependent optical properties of the grains, which lead to different occultation opacities in the visible and near-infrared, may provide direct constraints on grain properties \cite{bohren2008absorption}, as it has been proposed for Haumea's ring system \cite{Kalup2024}. 
  %{\color{blue} This we do not really use in the paper, and therefore this part, starting with 'Yet,...' is rather misleading! } --- rephrased suggestion above by Csilla

\begin{figure}[!ht]
\centering
\includegraphics[width=1.0\textwidth]{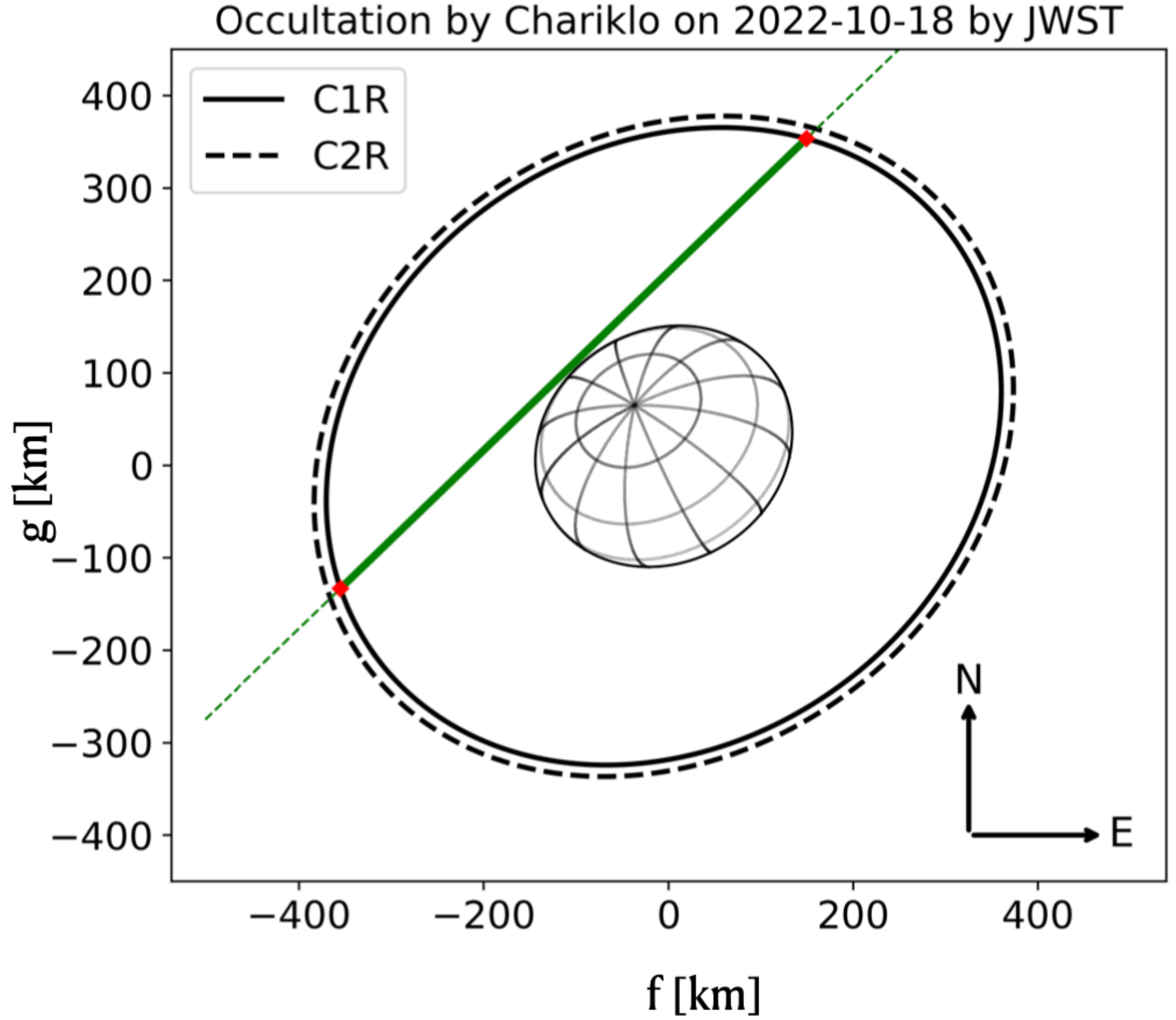}
\refstepcounter{extendeddatafig}
\caption*{\textbf{\theextendeddatafig:} Occultation map derived from the reconstructed JWST ephemeris. The map displays the relative positions of Chariklo’s main body and ring system at the time of occultation, based on the reconstructed JWST trajectory. The actual chord observed by JWST is also shown, revealing that the star crossed the rings without being occulted by the body itself, enabling a clear detection of the ring features.}
\label{fig:jwst_orbit_chariklo_prediction_1}
\end{figure}

 To investigate the fine structure and composition of Chariklo’s rings beyond the limitations of ground-based observations, we targeted a stellar occultation using the James Webb Space Telescope (JWST). Predicting such an event from a space-based observatory is inherently challenging (see Section \ref{subsec:predictions} for details); nevertheless, we successfully predicted and observed a stellar occultation by Chariklo on 18 October 2022. Remarkably, the JWST chord passed just $\sim$7.4 ~km from the body's centre (\ref{fig:jwst_orbit_chariklo_prediction_1}), allowing us to detect the ring system (see also Section~\ref{subsec:analysisrings}). No additional dips in brightness were observed that would indicate the presence of other material or orbiting satellites. This observation was conducted simultaneously in two near-infrared bands, centred at 1.5 and 3.2 $\mu$m (F150W2 and F322W2), including the first occultation detection ever beyond 3 $\mu$m, which is impossible from the ground. \textit{Exactly the same JWST setup} was used later in the observations of the occultation event by the ring system of Quaoar \citep{Proudfoot2025}. Although the radial locations of both Chariklo's rings in 2022 remain consistent with earlier measurements, they also exhibit significant variations compared to previous observations. 

For clarity, we adopt the terms \textit{first contact} ($1^{st}$ contact) and \textit{second contact} ($2^{nd}$ contact) to refer to the two points where the star’s line of sight intersects the ring plane, each comprising an ingress and an egress: Ingress$_1$–Egress$_1$ and Ingress$_2$–Egress$_2$, respectively. This terminology will be used throughout the paper.% {\color{blue} This is quite technical for the main text, and I do not think it is needed here. Moreover, I do not see how we 'cross' the ring plane??? We >>intersect<< the ring! }

\begin{figure}[!ht]
\centering
\includegraphics[width=\linewidth]{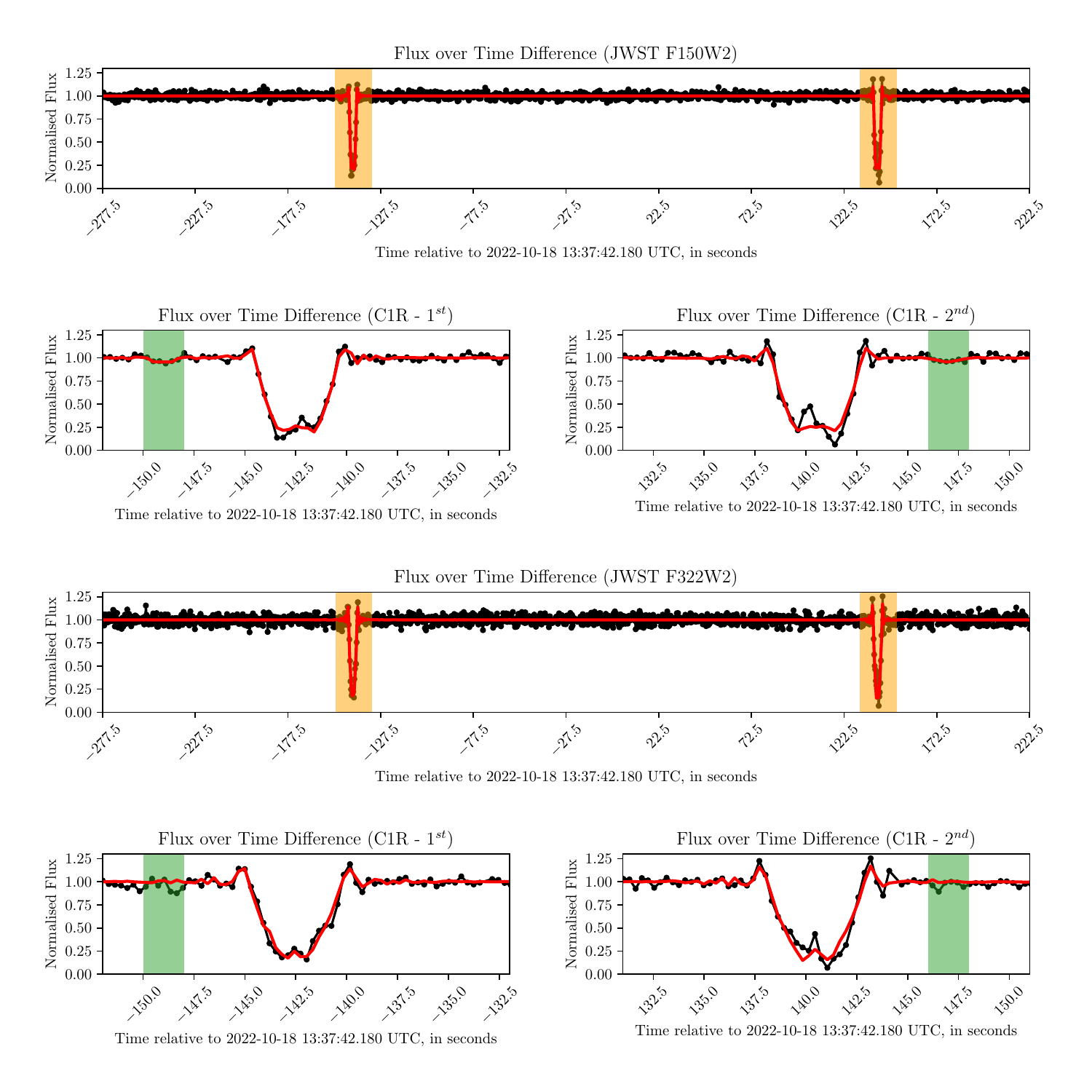}   
\caption{Light curves obtained with the James Webb Space Telescope (JWST) Near Infrared Camera (NIRCam), using the F150W2 (1.5 $\mu$m) and F322W2 (3.2 $\mu$m) filters, reveal the characteristic drop in stellar brightness as Chariklo’s rings occulted a background star on 18 October 2022 (UT). Although Chariklo’s body did not occult the star from JWST’s perspective, its rings did. In the upper panels, the locations of the C1R and C2R ring detections in both filters are highlighted in orange. The zoomed-in lower panels further illustrate these events, with clear diffraction spikes marking the sharp ingress and egress edges of the C1R ring. The light green shaded region indicates where C2R is detected or expected. Black dots represent the observed flux, while the red curve shows the best-fit model to the ring occultations.}
\label{fig:jwst_lcs}
\end{figure}

Chariklo’s inner dense ring, C1R, is clearly detected in the JWST light curves, exhibiting pronounced diffraction spikes at ingress and egress (see Figure \ref{fig:jwst_lcs}), indicative of sharp edges and strong confinement, morphological features reminiscent of Uranus’s rings \cite{Elliot1977, Elliot1987}.

Interestingly, C1R is found to be more opaque than before—with all uncertainties throughout this paper reported at the 1-$\sigma$ level—the pre-JWST measurements provide an average normal opacity of \( p_{N} = 0.303 \pm 0.028 \) for this ring, while considering only the JWST measurements gives \( p_{N} = 0.431 \pm 0.012 \), indicating a highly significant difference, corresponding to a z-score of z\,=\,4.2 and probability of \( p \approx 3 \times 10^{-5} \). %\textcolor{red}{, 
%that the JWST opacities are compatible with those of the previous, visible range occultation measurements}.
In contrast, the outer ring (C2R) shows an unexpectedly weak signal in JWST measurements. After applying an occultation model with a monolayer abrupt edge, including the effects of Fresnel diffraction (see Section \ref{subsec:analysisrings}), C2R was marginally detected at $\geq$1$\sigma$ in the F150W2 data, but not in the F322W2 filter, despite both measurements being performed simultaneously and sampling the same ring segment. Compared with previous ground-based visible-range observations, this suggests particularly low opacities for C2R. 

To unravel the origin of these differing opacities observed across epochs and wavelengths, we explore three basic scenarios that could explain these phenomena: 1) Azimuthally variable rings, where JWST consistently sampled a sparser or denser segment compared to earlier measurements; 2) Changes in the ring material, involving an increase or decrease in the total amount of material and/or the effective cross-section of grains, but without wavelength dependent grain properties;
3) Grains with wavelength-dependent optical properties, where composition and grain size effects are responsible for the observed variations.

In the case of a fragmented ring, alternating low- and high-density segments produce azimuthal variability in the observed ring structure. This affects both opacity and equivalent width ($E_P$), the latter of which combines opacity, radial width, and the instrument's sensitivity and time resolution (see Section \ref{subsec:analysisrings} for more information). While previous occultation measurements yield roughly consistent values for these parameters, the JWST observations reveal significantly different opacities and equivalent widths. Using a simple azimuthal ring density model, we estimate that the probability of randomly sampling a pair of (both ingress and egress) high-density (C1R) or low-density (C2R) values in the JWST observations against eight other events is very low, with p$\approx$0.004 for C1R and p$\lesssim$0.002 for C2R. See a comprehensive description of this model and calculation in Section \ref{subsec:ringsampling}.

\begin{figure}[H]
\centering

% Panel (a)
\begin{subfigure}[t]{\textwidth}
    \centering
    \includegraphics[width=\linewidth]{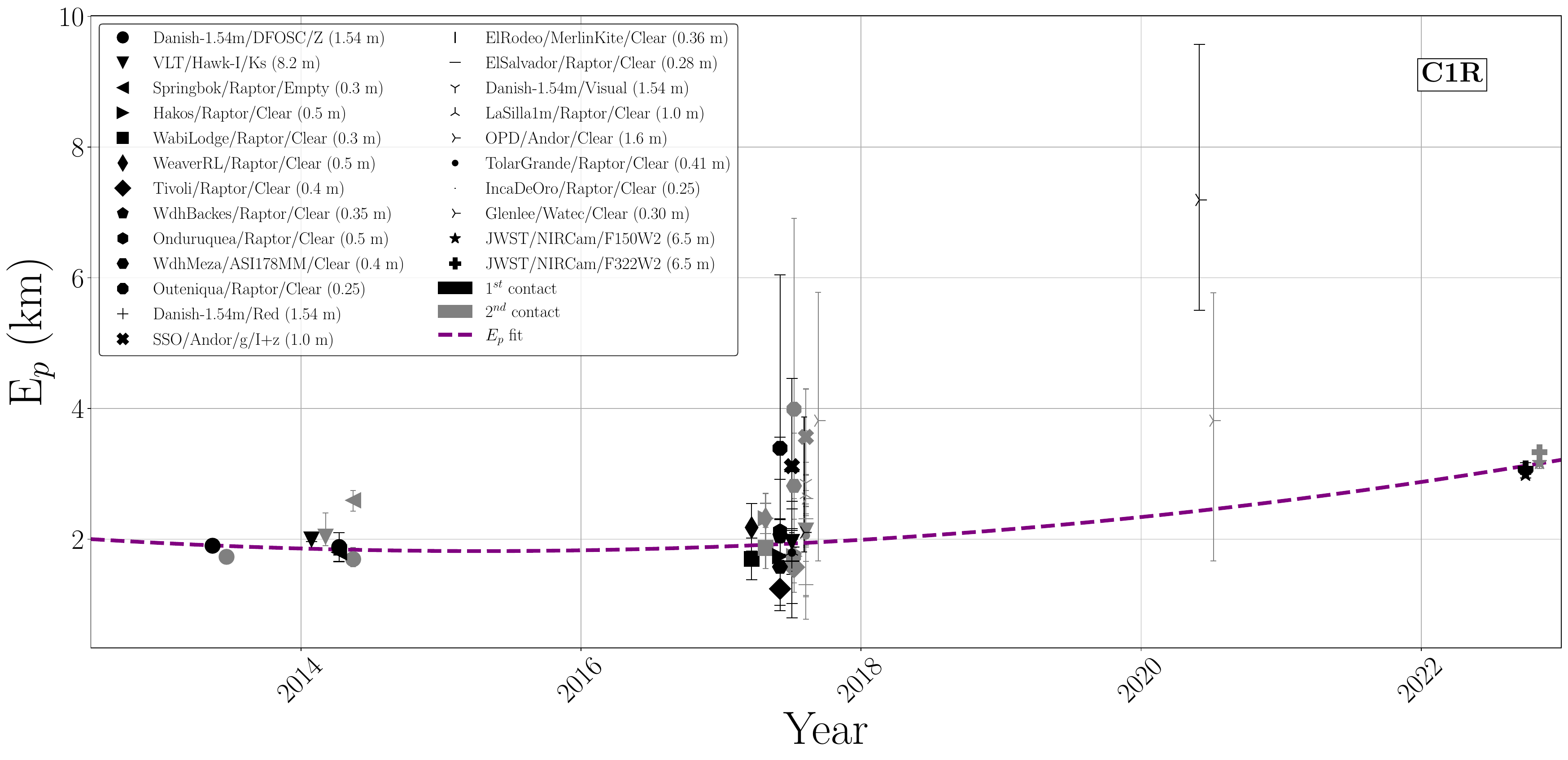}
    \caption*{(a)}
\end{subfigure}
\hfill

% Panel (b)
\begin{subfigure}[t]{\textwidth}
    \centering
    \includegraphics[width=\linewidth]{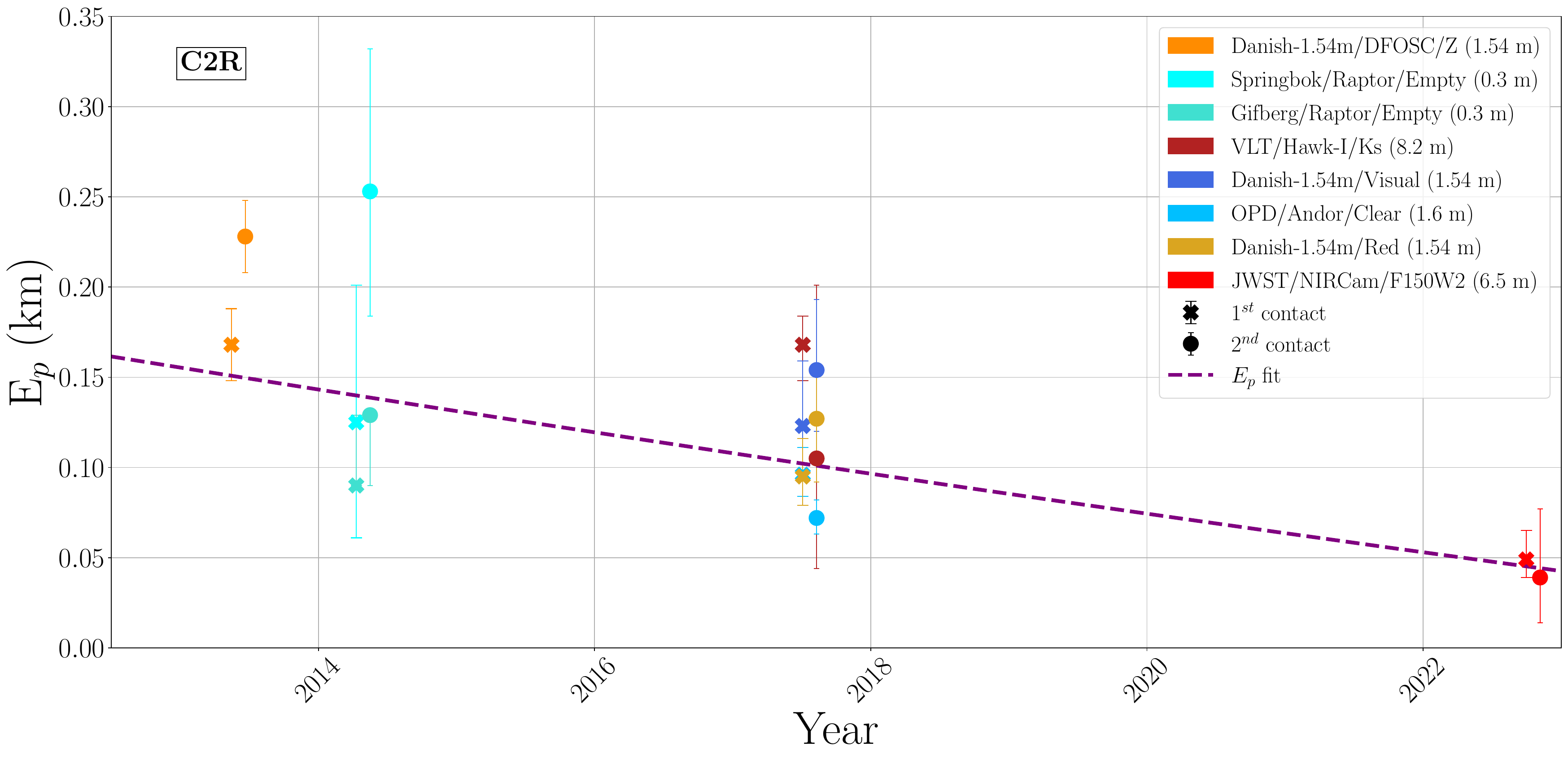}
    \caption*{(b)}
\end{subfigure}

\caption{Temporal evolution of Chariklo’s rings.
(a) Temporal evolution of the equivalent widths ($E_P$) of the C1R ring, derived from stellar occultation events between January 2013 and October 2022. Each marker corresponds to an individual $E_P$ measurement (in km), with asymmetric error bars reflecting observational uncertainties (see \ref{tab:c1r_equivalent_widths}). Marker shapes indicate the observing sites and instruments, while black and grey colours represent the 1$^{\text{st}}$ and 2$^{\text{nd}}$ ring contacts, respectively. A quadratic polynomial fit, obtained via the RANSAC (Random Sample Consensus) algorithm, is overplotted to highlight potential long-term variations or structural evolution within C1R.
(b) Time series of C2R equivalent width ($E_P$) measurements derived from multiple stellar occultation events spanning 2013–2022, illustrating a possible decline in opacity over the past decade. Each data point corresponds to a single $E_P$ estimate (in km), with asymmetric error bars indicating uncertainties (see \ref{tab:c2r_equivalent_widths}). Marker shapes and colours identify the observing sites and instruments, while ‘X’ and ‘o’ symbols represent the 1$^{\text{st}}$ and 2$^{\text{nd}}$ ring contacts, respectively. A quadratic polynomial fit, obtained via the RANSAC algorithm, is overplotted to emphasise the long-term trend of the ring.}

\label{fig:ep_over_time_c1r_c2r}
\end{figure}

Changes in the effective cross-section in occultation measurement are best described by the just-cited equivalent width ($E_P$). $E_P$ is proportional to the integrated effective cross-section of the grains, responsible for the extinction within the occulted segment (see Section \ref{subsec:analysisrings} for its formal definition). Combining the 2022 JWST occultation measurements with available ground-based data from 2013, 2014, and 2017 (see \ref{tab:c1r_equivalent_widths}, \ref{tab:c2r_equivalent_widths} and Section \ref{subsec:analysisrings} for details), we carried out a detailed study of the temporal evolution of the rings. In the case of C2R, while no clear trend is seen in $E_P$ between 2013 and 2017, a marked decrease ($\sim60\%$) is observed between 2017 and 2022 (Figure~\ref{fig:ep_over_time_c1r_c2r}, Panel (b)), potentially revealing a fast evolution in a timescale of a few years. Strikingly, the JWST opacity and equivalent width of C1R shows an opposite trend, a considerable increase ($\sim$50\%, see Panel (a) in Figure \ref{fig:ep_over_time_c1r_c2r}, \ref{tab:c1r_equivalent_widths}, and also \citep{Berard2017, Morgado2021}). This change in $E_P$ in C1R ($\Delta E_P^{C1R}=0.90$) is approximately 10 times greater than that in C2R ($\Delta E_P^{C2R}=0.094$), indicating that the material missing from C2R cannot account for the increase in the normal opacity and $E_P$ in C1R. This increase could either be due to an overall increase in the amount of material or to the rise in the effective cross-section of grains, for example, through collisions between larger particles or rocks inside the ring, producing a cascade of smaller grains.
The apparent fading of C2R may indicate that it is actively dispersing. Such depletion could result from collisions among particles, producing smaller grains that are more easily lost. Without effective confinement, these grains are subject to increasing orbital eccentricities due to radiation pressure, leading their orbits to intersect the inner ring \cite{Csaba2024}. Several known dusty rings in the Solar System exhibit temporal variability in opacity, morphology, or confinement, such as Saturn’s D, E, and F rings, Neptune’s Adams arcs, and Uranus’s $\lambda$ ring. For instance, Saturn’s D ring has shown measurable radial shrinkage and brightness changes on decadal timescales, likely driven by micrometeoroid bombardment or electromagnetic forces \cite{HEDMAN200789}. The E ring, sourced by Enceladus’s plumes, varies with the moon’s activity  and the surrounding magnetospheric environment. \cite{KEMPF2008420}. Neptune’s Adams arcs exhibit pronounced morphological evolution and brightness changes over months to years, suggesting a dynamically evolving configuration \cite{renner2014}. Uranus’s $\lambda$ ring also shows opacity changes possibly linked to variable confinement mechanisms \cite{FRENCH2012181}. These examples illustrate that dusty, low-mass rings can undergo rapid and significant evolution, reinforcing the interpretation of Chariklo’s outer ring as a transient, evolving structure. 

\begin{figure}
    \centering
    \includegraphics[width=1.\linewidth]{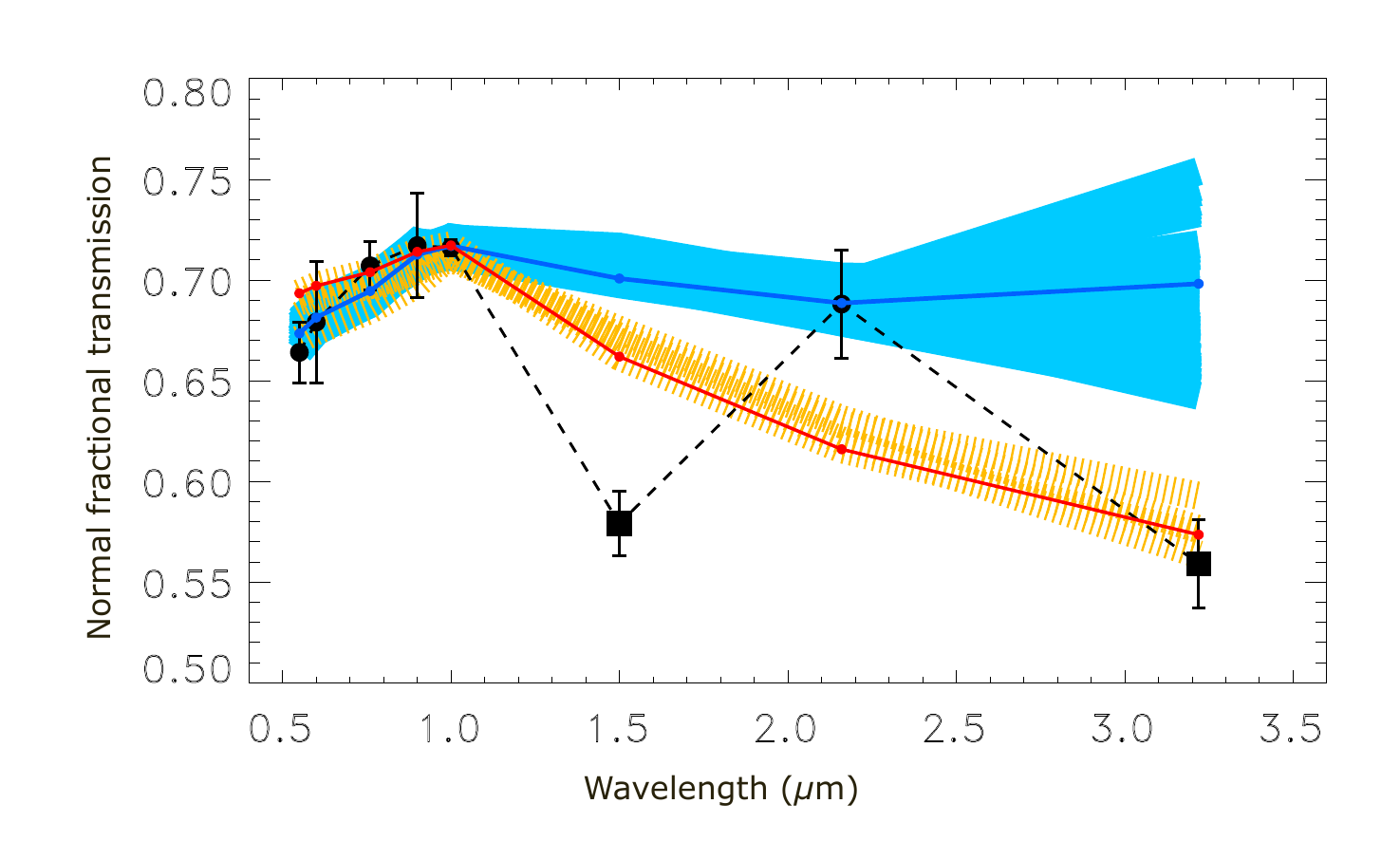}
    \caption{
    Comparison of observed normal fractional transmission ($1 - p_N$) with radiative-transfer models using different compositions and grain sizes for C1R. Black symbols represent the observed normal fractional transmission values (filled circles: pre-JWST observations, squares: JWST data). The solid blue curve shows the best-fit model (lowest $\chi^2$) using only pre-JWST measurements, while the solid red curve represents the best-fit model including the JWST data. The shaded blue (filled) and orange (striped) regions indicate the approximate parameter space where the models provide the best fits.
    }
    \label{fig:c1r_rad_trans}
\end{figure}

Since JWST probes a wavelength range where stellar occultations have rarely been observed, the discrepancy with visible-light data may also reflect the optical properties of the ring particles—specifically, a dominance of sub-micron grains whose scattering efficiency decreases at longer wavelengths. To assess whether these variations reflect a genuine wavelength dependence, we adopt a radiative transfer approach using various grain types (e.g., water ice, olivine, carbon) and particle sizes, and compare the resulting extinction values with available well-resolved occultation measurements (see \ref{tab:meas}) to provide a compatibility analysis of any plausible compositions that may be consistent with the observations. A detailed description of this modelling can be found in Section~\ref{subsec:radiative_transfer_modeling}. For C1R, we used a mixture of two materials with different compositions and potentially different grain sizes (see Section ~\ref{subsec:radiative_transfer_modeling} and Figure ~\ref{fig:c1r_rad_trans}). The pre-JWST data, including the 2017 VLT/Ks measurement at $\sim$2.1\,$\mu$m, can be well-fitted with mixtures dominated by micrometre-sized water ice or silicate grains with the contribution of tiny grains, also indicating a nearly constant opacity throughout the near-infrared. However, when JWST data are included, no solution (i.e., a combination of two materials and grain sizes) can simultaneously fit all datasets, particularly the JWST F150W2, F322W2, and VLT/Ks bands. The best-fit curves in this case consist predominantly of silicates with grain sizes of 1--10\,$\mu$m, but the overall fits are poor. This clearly indicates that the observed change in opacity cannot be solely attributed to wavelength dependence and leaves us with the most likely scenario: the opacities in C1R have indeed increased. For C2R, it is much harder to perform such an analysis due to the lack of well-resolved occultation profiles. However, as we introduced above, if C2R is actively depleting, this is a strong indication that small grains dominate this ring, as sub-micron grains may be inherently short-lived and particularly susceptible to dynamical processes that promote dispersal.

%{\color{ForestGreen} \sout{Wavelength invariance \textcolor{red}{can also} suggest a dust-poor population dominated by particles >1 mm, comparable to the centimetre-to-metre-sized grains in Saturn’s and Uranus’s dense rings \citep{Cuzzi2018, Nicholson2018}. Maintaining such structures likely involves the reversal of angular momentum flux driven by collisions among resonantly perturbed macroscopic particles \cite{Salo2024}. }}

%%%

However, small particle sizes raise questions about long-term stability. In general, radiation pressure could expel micron-sized grains on relatively short timescales, potentially leading to structural distortion or dispersal \citep{Csaba2024}. Yet, occultation data spanning nearly a decade reveal no significant changes in the radial position of both C1R and C2R (see Table~\ref{tab:occparams}), suggesting that a stabilising or replenishing mechanism must be active. The confinement of C1R is thought to result from the strong 1:3 spin-orbit resonance with Chariklo’s rotation, in which a ring particle completes one orbit for every three rotations of the central body \citep{Salo2024}. While effective for C1R, this resonance drives an outward migration of the ring, ultimately risking its departure from the resonance and subsequent dispersion. A small external shepherd moon has been hypothesised to prevent such drift and maintain C1R’s location \citep{Salo2024}. If present, this moonlet might share its orbit with C2R and act as a source of ring material, analogous to the small satellites observed in Saturn’s and Uranus’s ring systems. This scenario is further supported by our estimate of C2R’s dynamical spreading timescale: based on JWST measurements, we find a median value of only 0.34 years (1$\sigma$ range: 0.21–0.96 years; see Methods, Section~\ref{subsec:time_scale_of_c2r}), implying that without confinement or replenishment, the ring should have dispersed since its discovery in 2013. A co-orbiting shepherd satellite could thus play a dual role—maintaining the ring’s sharp edges while continuously supplying fresh material \citep{Sickafoose2024}. However, another recent dynamical study about the effect of solar radiation pressure on the stability of a small-body rings shows that for highly tilted rings, such as the one around Chariklo, there are two plausible particle size ranges that can be stable: $\lesssim2.5-15\,\mu$m and $\lesssim60-300\,\mu$m \citep{Regaly2025}. If the outer ring is indeed degrading over a timescale of just a few years, this would mark it as a transient structure, potentially representative of a broader class of ephemeral rings among small bodies, as hinted by episodic activity and temporarily bound structures around another Centaur, Chiron \citep{Ortiz2023}.

%¿Cuánto tiempo viven los granos pequeños y de dónde vienen?
%The emerging picture suggests that the two scenarios for C2R—rapid erosion and a dusty, micron-scale particle population—are not mutually exclusive. A ring composed of sub-micron grains may be inherently short-lived, as such particles are particularly susceptible to dynamical processes that promote dispersal. We therefore consider both effects jointly, as they may together account for the observed decline in opacity and signal strength of the C2R ring. If instead the C2R ring is linked to Chariklo’s activity as a Centaur-like object, we would again expect small grains, as cometary dust is predominantly sub-micron in size \citep{Wooden2017, Harker2023}. Crucially, a single occultation in visible light of a sufficiently bright star could distinguish between these scenarios: if the opacity matches values from 2013–2017, grain-size effects are favoured; but if the occultation yields significantly lower opacities or no detection at all, it would instead support the interpretation of an actively fading ring. Either outcome would raise fundamental questions about the longevity of dusty rings and the fate of metre-sized material subject to slow but persistent erosion over decadal timescales. 

A single near-future occultation measurement in visible light (different from JWST wavelengths) could confirm the fading of C2R and the strengthening of C1R. Ultimately, our JWST observations unveil a new and unexpected level of complexity in minor body ring systems—one that challenges existing paradigms and opens an unprecedented window into the transient architectures sculpting the outer Solar System. These findings provide the first evidence of such episodic changes in these seemingly frequent systems, revealing a dynamical behaviour that had remained entirely hidden until now.
%Ultimately, our JWST observations unveil a new and unexpected level of complexity in minor body ring systems—one that challenges existing paradigms and offers an unprecedented window into transient processes. These observations provide the first evidence of such episodic changes in these seemingly frequent systems in the outer Solar System, which had remained entirely hidden until now.

\begin{sidewaystable}[htbp]
    \centering
    \small
    \caption{Fitted Chariklo ring parameters obtained from the JWST stellar occultation observation.}
    \renewcommand{\arraystretch}{1.5}
    \begin{tabular}{c c c c c c c} \hline \hline
    \multirow{3}{*}{Parameter}      & \multicolumn{4}{c}{F150W2}                        & \multicolumn{2}{c}{F322W2}                        \\
                                    & \multicolumn{2}{c}{C1R} & \multicolumn{2}{c}{C2R} & \multicolumn{2}{c}{C1R} \\
                                    & 1$^{st}$ contact               & 2$^{nd}$ contact               & 1$^{st}$ contact               & 2$^{nd}$ contact               & 1$^{st}$ contact               & 2$^{nd}$ contact                  \\  \hline
    $T$                             & $0.24\pm0.01$        & $0.25\pm0.02$        & $0.887^{+0.922}_{-0.738}$        & $0.960^{+0.974}_{-0.950}$         & $0.23\pm0.02$        & $0.22\pm0.02$          \\
    Mid-time                        & 13:35:19.74$\pm0.04$ & 13:40:02.72$\pm0.07$ & 13:35:13.15$\pm0.32$ & 13:40:09.15$\pm0.39$ & 13:35:19.64$\pm0.09$ & 13:40:02.66$\pm0.05$    \\
    $r$ [km]                        & $385.9\pm0.1$        & $385.9\pm0.2$        & $400.3\pm0.1$        & $400.7\pm0.4$        & $386.0\pm0.2$        & $385.8\pm0.2$           \\
    $W_r$ [km]                      & $7.04\pm0.07$        & $7.6\pm0.1$          & $1.009^{+1.492}_{-0.386}$          & $2.273^{+4.897}_{-0.862}$          & $7.0\pm0.2$          & $7.5\pm0.1$             \\
    $p'$                            & $0.76\pm0.01$        & $0.75\pm0.02$        & $0.113^{+0.262}_{-0.078}$        & $0.040^{+0.050}_{-0.026}$        & $0.77\pm0.02$        & $0.78\pm0.2$            \\
    $p_{\rm{N}}$                    & $0.425\pm0.008$        & $0.417\pm0.013$        & $0.048^{+0.117}_{-0.033}$        & $0.017^{+0.021}_{-0.011}$        & $0.438\pm0.017$        & $0.444\pm0.014$           \\
    $\tau'$                         & $1.42\pm0.04$        & $1.39\pm0.06$        & $0.120^{+0.304}_{-0.081}$        & $0.041^{+0.052}_{-0.026}$        & $1.49\pm0.09$        & $1.52\pm0.07$           \\
    $\tau_{\rm{N}}$                 & $0.59\pm0.02$        & $0.58\pm0.03$        & $0.050^{+0.127}_{-0.034}$        & $0.017^{+0.022}_{-0.011}$        & $0.62\pm0.04$        & $0.63\pm0.03$           \\ 
    $E_{\rm{p}}$ [km]               & $2.99\pm0.05$        & $3.17\pm0.09$        & $0.049^{+0.065}_{-0.039}$        & $0.039^{+0.077}_{-0.014}$        & $3.08\pm0.09$        & $3.33\pm0.10$           \\ 
    $A_{\tau}$ [km]                 &  $4.2\pm0.1$         &  $4.4\pm0.2$         & $0.050^{+0.067}_{-0.040}$        & $0.039^{+0.078}_{-0.014}$        & $4.4\pm0.2$         &  $4.8\pm0.2$            \\ \hline
    \end{tabular}
    \label{tab:occparams}
    \textit{Note:} $T$: Transmittance; Mid-time: midpoint time of each occultation contact; $r$: radial distance in the ring plane (km); $W_r$: radial width in the ring plane (km); $p'$: apparent opacity; $p_{\rm{N}}$: normal opacity; $\tau'$: apparent optical depth; $\tau_{\rm{N}}$: normal optical depth; $E_{\rm{p}}$: equivalent width (km); $A_{\tau}$: equivalent depth (km). C2R was not detected in the F322W2 filter due to a low signal-to-noise ratio. 1$^{st}$ and 2$^{nd}$ contacts refer to the ones before and after the closest approach, respectively. Values are presented with uncertainties: asymmetric uncertainties are indicated with upper and lower limits ($^{+\text{upper}}_{-\text{lower}}$), while symmetric uncertainties are shown with a single value following the $\pm$ symbol.
\end{sidewaystable}
\newpage

\section{Methods}
\label{sec:methods}

\subsection{Predicting and observing stellar occultations from space}
\label{subsec:predictions}

In June 2020, the first-ever predicted stellar occultation observation using a space telescope was obtained with the Characterising ExOPlanets Satellite (CHEOPS) \cite{Morgado2022}. This observation demonstrated the feasibility of predicting and observing stellar occultations by distant solar system minor bodies using space telescopes: these facilities are unaffected by weather and can extend the range of observations beyond Earth's surface. The use of space telescopes also facilitates the detection of material surrounding small bodies, enabling photometric accuracy comparable to that achieved by ground-based telescopes with much larger apertures. We also have access to stellar occultations of fainter stars than we do from the ground, thereby multiplying the number of possible events.

\begin{figure}[ht]
\centering
\includegraphics[width=\textwidth]{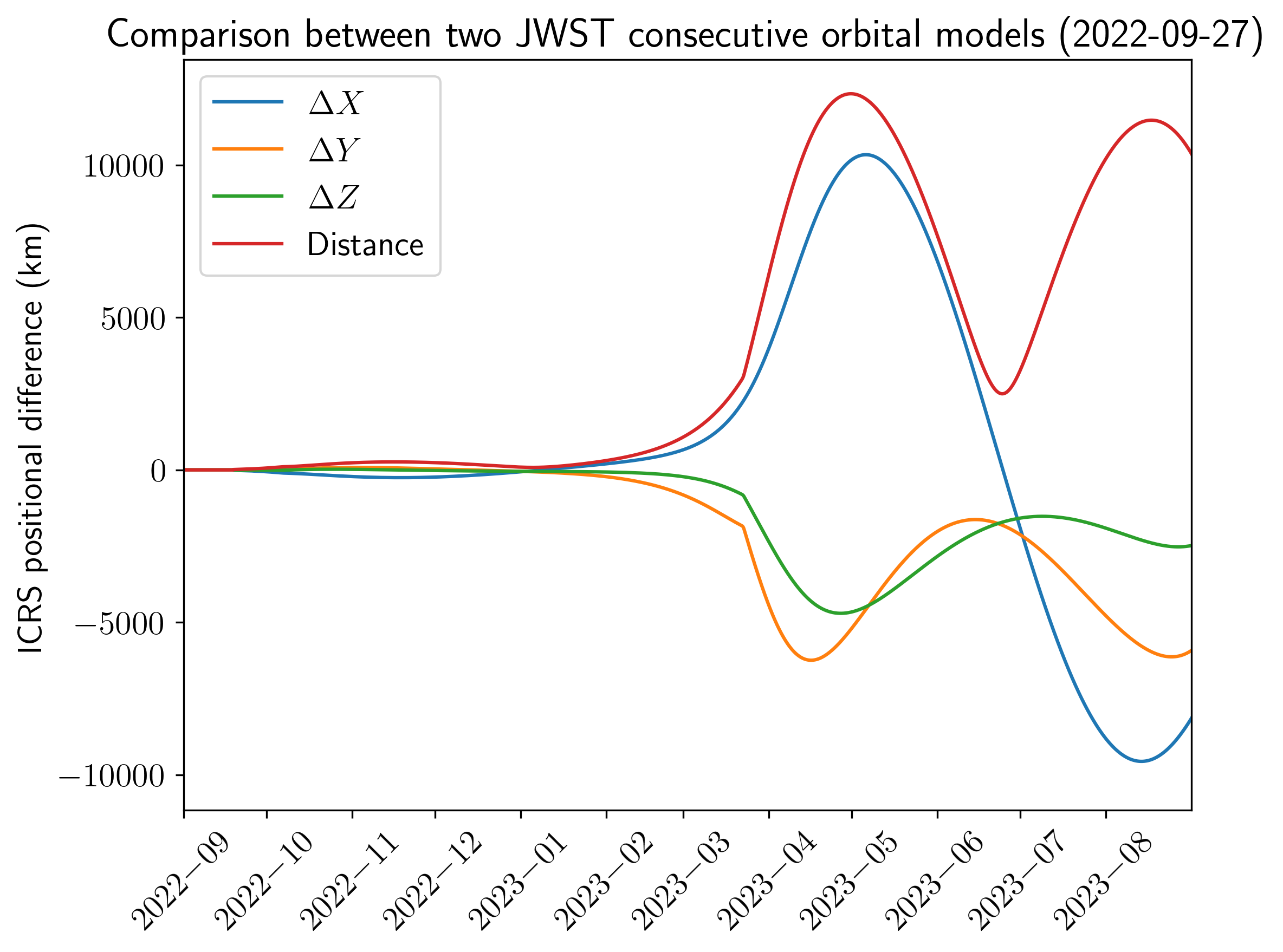}
\refstepcounter{extendeddatafig}
\label{fig:jwst_ephemeris}
\caption*{\textbf{\theextendeddatafig:} Positional discrepancy between successive JWST ephemeris solutions. The plot shows the difference in Chariklo’s predicted position using two consecutive JWST ephemeris releases, both issued on 27 September 2022. The offset highlights the sensitivity of occultation predictions to small updates in spacecraft trajectory data, underscoring the challenge of planning such events from a moving observatory in L2 orbit.
}
\end{figure}

JWST is orbiting close to the second Sun-Earth Lagrange point (L2) \cite{Gardner2023, McElwain2023}. L2 is an unstable equilibrium point, meaning constant manoeuvres must be performed to keep the telescope in orbit around it \cite{Menzel2023}: these are usually performed 2--3 times a month. Consequently, the predicted orbital path of JWST needs to be frequently updated (J. Giorgini, personal communication, June 22, 2022). For all the above reasons, and given the complex motion of the space telescope around L2, predicting stellar occultations by distant minor bodies visible from JWST is a very challenging task. \ref{fig:jwst_ephemeris} shows the difference between two consecutive releases of the JWST ephemeris, obtained on September 27, 2022. In just under six months from the initial epoch, the difference between the two releases reached approximately 10,000 km, exceeding the Earth's radius. In the absence of formal positional uncertainty, the plot illustrates the trustworthiness of a stellar occultation prediction using JWST far from the initial epoch.

To predict stellar occultations that JWST could potentially observe, we used the Stellar Occultation Reduction and Analysis (SORA) software package \cite{GomesJunior2022} combined with the \textit{Gaia} DR3 catalogue \cite{Gaia2016, Gaia2023} and the NIMA ephemeris \cite{Desmars2015} for the solar system bodies. The strategy we developed to run the predictions is as follows. We selected the 40 largest TNOs/Centaurs (36 TNOs and 4 Centaurs), then ran predictions for these 40 targets every week for \textit{Gaia} DR3 stars up to a G-magnitude of 20. This re-running of the predictions was necessary because of the station-keeping manoeuvres applied to JWST, which changed its position (ephemeris). Each week, we reviewed the updated predictions to identify favourable events, refined them with new ground-based astrometric data, and reran the predictions for those objects. Finally, when we were confident in the prediction, we activated the target of opportunity (ToO) to observe the occultation from JWST. Since this was a non-disruptive ToO, it had to be activated at least 14 days before the event. 

\begin{figure}[ht]
\centering
\includegraphics[width=\textwidth]{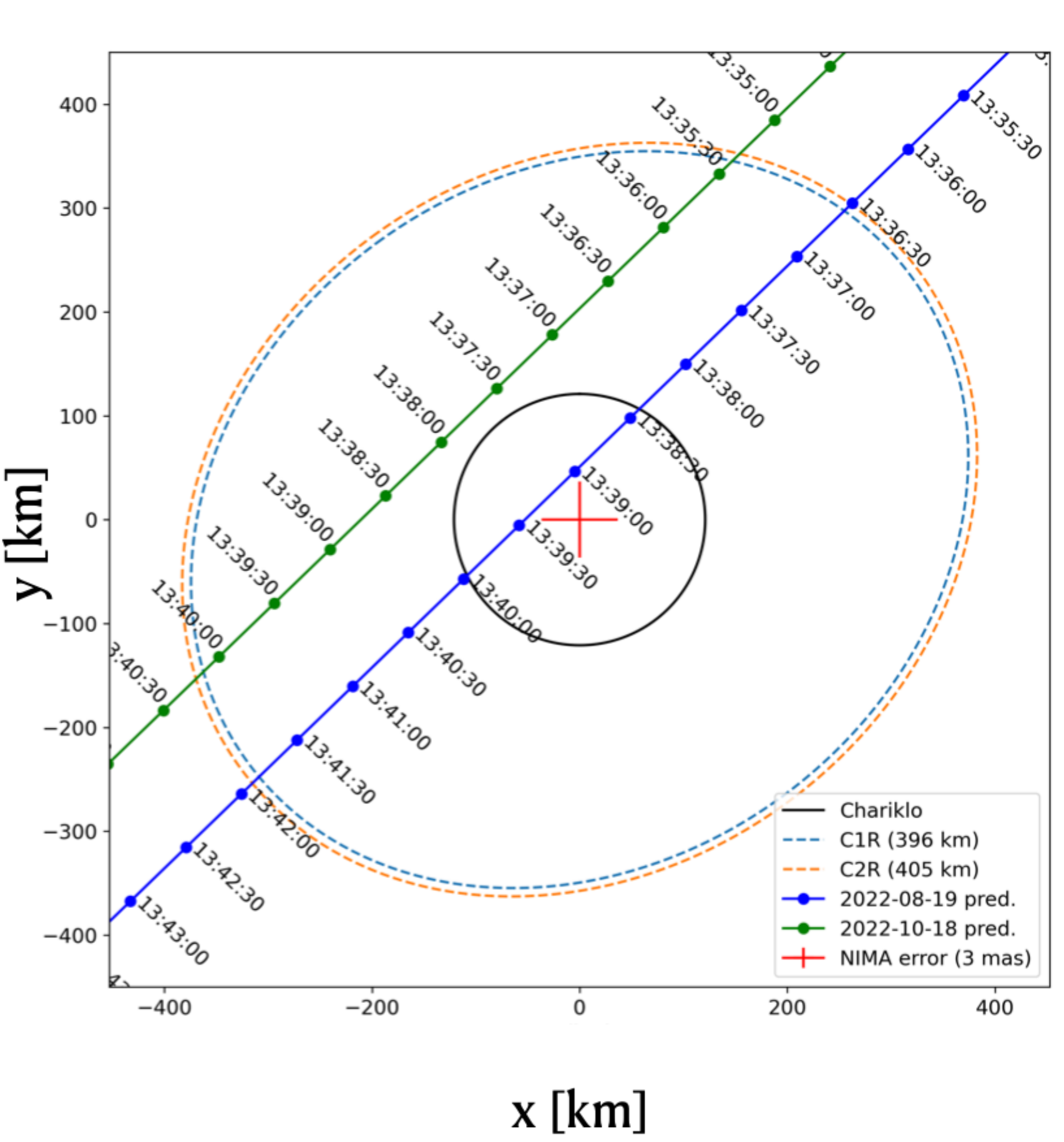}
\refstepcounter{extendeddatafig}
\label{fig:jwst_orbit_chariklo_prediction_2}
\caption*{\textbf{\theextendeddatafig:} Refinement of Chariklo occultation predictions. Occultation prediction maps derived from JWST ephemerides issued on 19 August 2022 (blue line) and a few hours before the event on 18 October 2022 (green line). The comparison illustrates the improvement in accuracy enabled by last-minute orbital updates, critical for the successful observation of such a finely timed event.}
\end{figure}

Using the described strategy, we identified the occultation by the ringed Centaur Chariklo on 18 October 2022 (UT), with a low relative velocity (2.5 km/s) and a star bright enough to be observed by JWST (K = 15.7 mag) as the most promising event. We first identified this occultation on 19 August 2022 (blue line in \ref{fig:jwst_orbit_chariklo_prediction_2}). Then we reran the prediction each week to account for changes in JWST's orbit caused by station-keeping manoeuvres. Our most recent prediction was made on the day of the occultation, just a few hours before the event (green line in \ref{fig:jwst_orbit_chariklo_prediction_2}). In just about two months, the path of the star behind Chariklo, as seen by JWST, moved 110 km, enough to miss the occultation by the main body. Furthermore, the instant of occultation also shifted by 1.5 minutes.

\subsection{Observations}
\label{subsec:observations}

We used the James Webb Space Telescope (JWST) to observe a stellar occultation by Chariklo’s rings, marking the first time such an event was both predicted and successfully observed by JWST. It also represents the first stellar occultation by a small Solar System body ever recorded from beyond Earth’s orbit, with the only comparable case being a previous Quaoar occultation observed by the Earth-orbiting CHEOPS satellite. This observation was both scientifically valuable and technically demanding. It was conducted under JWST’s Solar System Guaranteed Time Observations (GTO) program \#1271, \textit{ToO TNOs: Unveiling the Kuiper belt by stellar occultations} (PI: Pablo Santos-Sanz) \cite{SantosSanz2017}.

On 18 October 2022, between 12:36:04 and 13:50:27 UT, we observed the event with the JWST NIRCam instrument, targeting the star \textit{Gaia} DR3 6873519665992128512 (RA $\approx$ 20$^h$ 21$^m$ 46$^s$, Dec $\approx$ -16$^{\circ}$ 26$'$ 22$''$; G${mag}$ = 17.5, K${mag}$ = 15.7). Observations were performed simultaneously using the wide F150W2 (0.6–2.3 $\mu$m) and F322W2 (2.4–5.0 $\mu$m) filters to maximise the stellar flux, a key factor given the high-cadence time series required.

We used the smallest available NIRCam subarray, SUB64P (64 $\times$ 64 pixels; 2$''$ $\times$ 2$''$ and 4$''$ $\times$ 4$''$ fields of view for the short- and long-wavelength channels, respectively), in combination with the RAPID readout pattern. Each integration comprised five non-destructive reads (groups), with a total exposure time of 0.304 s and a dead time of only 0.0009 s between integrations, resulting in a cadence of $\sim$3.3 Hz. At Chariklo’s distance ($\sim$17 au), this cadence corresponds to a spatial resolution of $\sim$750 m per integration. In total, 14,827 integrations were acquired over the 1.25-hour observation window. All data were processed using JWST calibration pipeline version 1.10.0 and the reference file context \textit{jwst\_1075.pmap}.

\subsection{Photometry: occultation light curves}
\label{subsec:photometry}

\begin{figure}[ht]
\centering
\includegraphics[width=\linewidth]{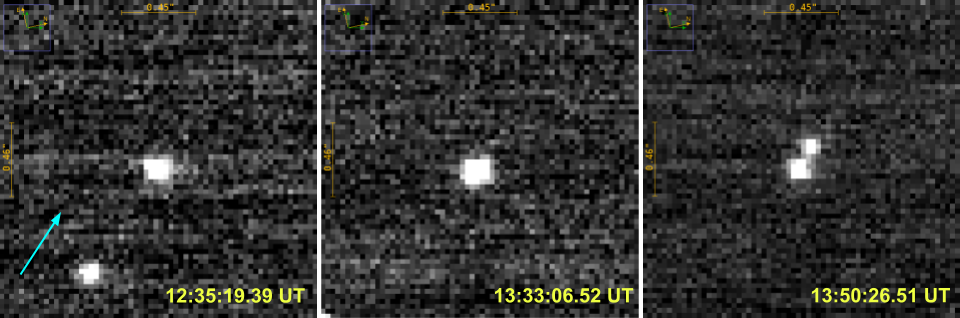}
\refstepcounter{extendeddatafig}
\label{fig:3example_images_F150W2}%
\caption*{\textbf{\theextendeddatafig:} NIRCam images during the Chariklo occultation in the F150W2 band. Three selected frames illustrate the relative positions of Chariklo and the occulted star. Left panel: Before the occultation, Chariklo (lower-left source) and the target star (central source) are clearly separated. Middle panel: During the occultation, both sources appear merged. Right panel: Following the event, the star (central source) and Chariklo (offset to the upper right) are again resolved. The mean time of each frame is indicated. The light blue arrow in the left panel marks Chariklo’s direction of motion.}
\end{figure}

Aperture and point spread function (PSF) fitting photometry were performed independently on the JWST/NIRCam time-series data obtained with the F150W2 and F322W2 filters. Aperture photometry yielded a lower dispersion than PSF-fitting methods and was thus adopted for the occultation analysis. Given the minimal field of view (FoV) of the SUB64P subarray, no suitable reference stars were available for differential photometry; only Chariklo and the occulted star were present in the images (see left Panel in \ref{fig:3example_images_F150W2}). The aperture radius was optimised to maximise the signal-to-noise ratio (SNR) of the combined flux from Chariklo and the star, while minimising photometric scatter. Fluxes were normalised to unity outside the occultation events.

The analysis focused on a subset of 1,646 images (out of 14,827 total exposures), centred around the predicted occultation time. The measured flux dispersion was 0.025 in the F150W2 filter and 0.041 in F322W2. A small fraction of the images—approximately 2\% in F150W2 and 1.2\% in F322W2—were affected by unidentified artefacts and excluded from the analysis.

The resulting light curves (Figure \ref{fig:jwst_lcs}) display the drop in stellar flux as Chariklo’s rings transited the background star. As anticipated from the occultation geometry (\ref{fig:jwst_orbit_chariklo_prediction_2}), the main body of Chariklo did not intercept the line of sight, and no additional dips beyond the 2$\sigma$ noise level were detected, even when the complete data set was examined.

Fluxes from Chariklo and the target star were also measured separately in both filters, using images acquired before the occultation when the two sources were spatially resolved (\ref{fig:3example_images_F150W2}, left Panel). These measurements were used to calibrate the relative contributions of Chariklo and the star to the combined flux, a crucial step in deriving the ring parameters discussed in Section \ref{subsec:analysisrings}. The fractional flux contribution of Chariklo (relative to the combined Chariklo + star flux) was 0.2780 for F150W2 and 0.2662 for F322W2.

\subsection{Analysis of the rings}
\label{subsec:analysisrings}

In the obtained light curves, the main, densest, and most opaque ring of Chariklo is clearly visible. This ring is called C1R, following the original notation \cite{Braga-Ribas2014}. The C1R occultation light curve displays prominent diffraction spike features immediately before and after the ingress and egress, respectively. These spikes provide strong evidence that the ring has sharp edges and is tightly confined, similar to the rings of Uranus \cite{Elliot1977, Elliot1987}. However, the dynamical mechanism responsible for this confinement is still unknown and is currently under discussion \cite{Sicardy2019, Sicardy2020a, GiuliattiWinter2023, Salo2024}.

\begin{figure}[ht]
\centering
\includegraphics[width=\linewidth]{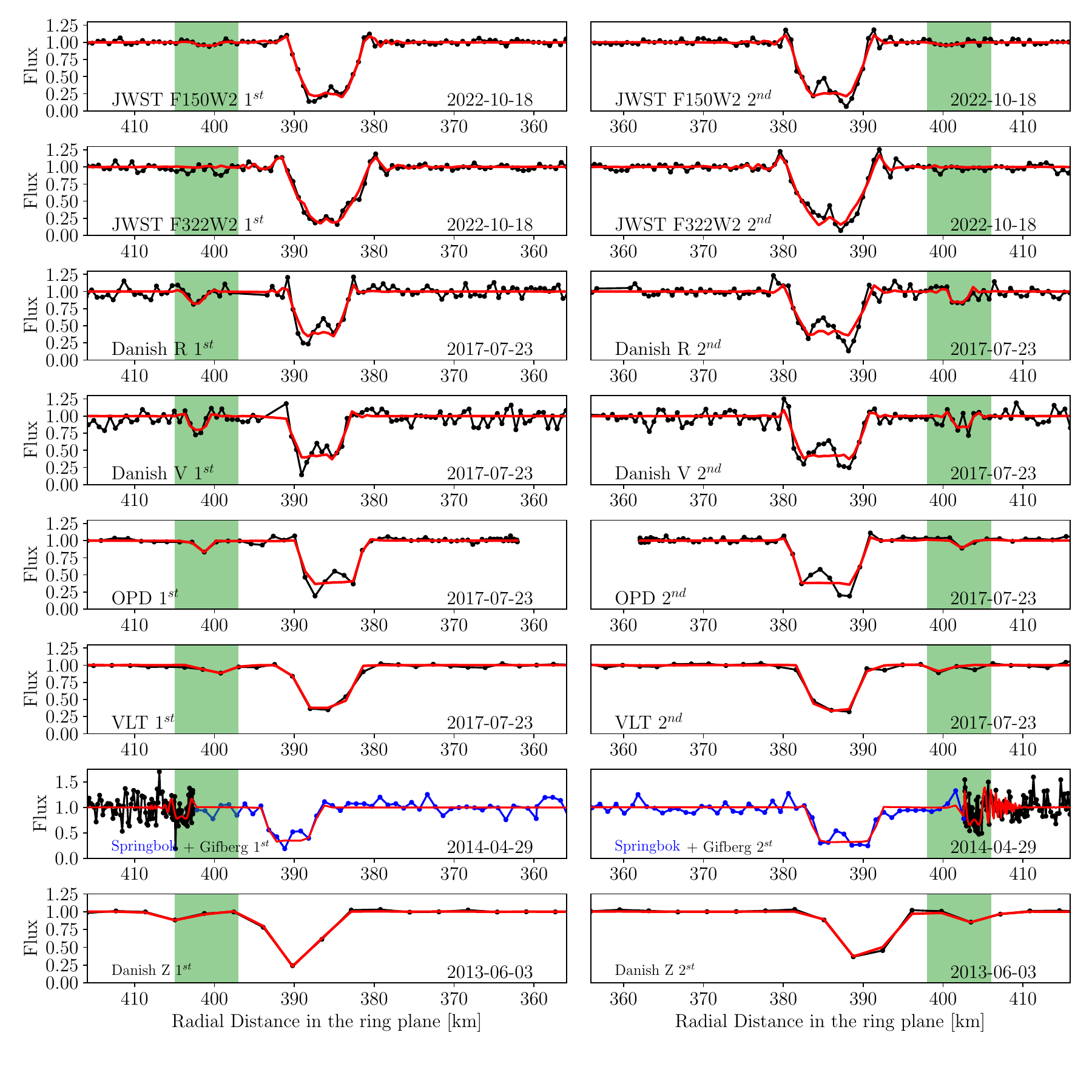}
\refstepcounter{extendeddatafig}
\caption*{\textbf{\theextendeddatafig:} Comparison between previous stellar occultations by Chariklo’s rings and the JWST/NIRCam light curves from 18 October 2022. Observed light curves are shown as black points (blue for the Springbok dataset), while red curves represent the best-fit ring models. The inner ring, C1R, is unambiguously detected in both NIRCam filters (F150W2 and F322W2). In contrast, the outer ring (C2R) is not clearly detected. The green shaded region marks the expected location of C2R in the JWST light curves, based on prior occultation geometry.}
\label{fig:LCs_comp}%
\end{figure}

The other ring (referred to as C2R) is surprisingly not clearly detected in the F150W2 or F322W2 light curves, as shown in \ref{fig:LCs_comp}. Since a ring partially occults the starlight, we fit an occultation model with a monolayer abrupt edge, including the effects of Fresnel diffraction, finite bandwidth, exposure time, and the \textit{Gaia} DR3 stellar diameter. We use the procedures within the SORA package \cite{GomesJunior2022}. From this analysis, we obtained the time when the star disappeared and reappeared behind Chariklo’s rings ($t_{i}$ and $t_{e}$) and the observed apparent ring opacity ($p'$), which measures the fractional drop of stellar flux.

Knowing the reconstructed JWST position at the time of the occultation, we were able to propagate the instants of disappearance and reappearance of the star into the tangent plane and determine two-dimensional coordinates ($f$, $g$) of the rings' positions relative to Chariklo's expected centre. This projection was executed using the SORA package, and its result is shown in Figure \ref{fig:jwst_lcs}. Assuming Chariklo's parameters (body and rings) as published by \cite{Morgado2021}, we fitted the centre position of Chariklo as viewed by JWST at the mid-point of the occultation (18 October 2022 at 13:37:42.18 UT), which resulted in
\begin{eqnarray}
RA &=&  \phantom{-}20^h~21^m~45^s.6825953~\pm~0.504~\textrm{mas}, \nonumber \\
DEC &=& -16^{\circ}~26'~21''.847603\phantom{0}~\pm~0.722~\textrm{mas}. \nonumber
\end{eqnarray}
That meant an appulse with the closest approach of 132.2 km, in contrast to Chariklo's volume-equivalent radius of 124.8 km \cite{Morgado2021}.

Initially, the parameters of both rings (C1R and C2R) were obtained in the sky plane simultaneously, just before and after the closest approach. We used the formalism presented by \cite{Berard2017} to project these parameters to the ring plane so we could obtain the relevant parameters, such as the ring radial width ($W_{\sf r}$), the normal opacity ($p_N$), and the normal optical depth ($\tau_N = -\ln(1 - p_N)$). 

Assuming a monolayer ring, the observed ring opacity normal to the ring ($p_{obs}$) can be described as
\begin{equation}
    p_{obs}=~|\sin(B)| p', \label{eq:normal_opacity_obs}
\end{equation}
alternatively, if we assume a poly layer ring, $\tau_{obs}$ is defined as \cite{Elliot1984}
\begin{equation}
    \tau_{obs}~=~|\sin(B)|\tau', \label{eq:normal_optical_depth_obs}
\end{equation}
where $B$ is the ring opening angle, and $B= 90^\circ$ (resp. $B= 0^\circ$) corresponds to a pole-on (resp. edge-on) viewing. On the other hand, considering that the Airy scale is larger than the width of the ring as seen in the sky plane \cite{Boissel2014}, a factor is needed to compute the fraction of lost light; in that case
\begin{eqnarray}
    p_N~&=&~|\sin(B)| \left( 1 - \sqrt{1 - p'} \right), \\ \label{eq:normal_opacity}
    \tau_N~&=&~|\sin(B)| \frac{\tau'}{2}, \label{eq:normal_optical_depth}
\end{eqnarray}
for a monolayer or polylayer ring, respectively.

The integrals of $p_N$ and $\tau_N$ over the radial width ($W_{\sf r}$) of the ring define the equivalent width ($E_p=~p_N W_{\sf r}$) and the equivalent depth ($A_{\tau}=~\tau_N W_{\sf r}$) of the profile. These values ($E_p$ and $A_{\tau}$) are proportional to the amount of material present in the profiles for each case \cite{Berard2017}. The ring parameters are obtained in Table \ref{tab:occparams}. We highlight that the detection of C2R is $\geq$1$\sigma$ significant in the F150W2 data and cannot be detected in the F322W2 data.
 
\subsection{Sampling a ring with azimuthally variable structure}
\label{subsec:ringsampling}

In the case of a ring with an azimuthally variable structure, we may observe different normal opacities ($p_N$) or equivalent with ($E_P$) values at the 1$^{\text{st}}$ and 2$^{\text{nd}}$ contacts during an occultation event, as well as variations between multiple occultations. If the ring opacity is assumed to be wavelength-independent, the observed decrease/increase in opacity and/or in $E_P$ during the 2022 \textit{JWST} events---relative to previous occultations---could be explained by the possibility that both ingress and egress sampled low/high-density segments of the ring.
To estimate the likelihood of such an occurrence, we performed a simple Monte Carlo simulation. We modelled an azimuthally variable ring, with the material distribution given by $p_N$ or $E_P$
\( \propto \sin^2(n\phi) \), 
where $\phi \in [0,2\pi]$ is the azimuthal angle and $n$ is a small integer. Random chords drawn through this ring yield a distribution of `observed' $p_N$ or $E_P$ values. 

In the real occultation measurements of the C2R ring, we obtain $E_P \geq 0.05$ in seven of the occultations (see \ref{tab:c2r_equivalent_widths}), with significantly lower values observed only during the \textit{JWST} event. For simplicity, we assume that the minimum of $E_P$ along the ring is zero, although a constant positive offset may exist. Therefore, the probability of randomly sampling a pair of such low values in just one of eight events, as calculated from the simulation, represents an \textit{upper limit}, with a probability of $p \leq 0.002$ obtained in our simulation.
Similarly, we obtain a probability of p\,$\approx$\,0.004, assuming that we sample the ring near the density maxima in the JWST measurements and below these values in the pre-JWST cases (\ref{tab:meas}). 
While azimuthal fragmentation is plausible---especially given the existence of fragmented rings in the trans-Neptunian region, e.g. in the case of Quaoar \citep{Morgado2022, Proudfoot2025}---these very low probabilities suggest that the observed \textit{JWST} opacity drop (C2R) or increase (C1R) is unlikely to be solely due to this effect. The results are independent of the parameter $n$ for $n$\,$\geq$\,3. Here, we assumed that the azimuthal structures of the two rings are independent.

\subsection{Spreading timescale of Chariklo’s C2R Ring}
\label{subsec:time_scale_of_c2r}
We compute the characteristic spreading timescale $T_S$ of Chariklo’s second ring (C2R) using the standard expression for a narrow, self-gravitating ring in resonance \cite{DePater2018}:
\begin{equation}
T_S = \frac{4\pi}{3\sqrt{GM}} \frac{a^{5/2}}{\delta a}
\label{eq:TS}
\end{equation}
where $G$ is the gravitational constant, $M$ the mass of Chariklo, $a$ the ring semi–major axis, and $\delta a$ its radial width. Chariklo’s mass is derived from its bulk density $\rho$ and ellipsoidal volume:
\begin{equation}
V = \frac{4}{3}\pi a_{\rm Ch} b_{\rm Ch} c_{\rm Ch},
\quad M = \rho V
\end{equation}
We adopted ellipsoidal semi-axes from Morgado et al. (2021)\cite{Morgado2021}: $a_{\rm Ch}=143.8\, $km, $b_{\rm Ch}=135.2\, $km and $c_{\rm Ch}=99.1\, $km and a mean bulk density of $\rho = 0.79~\mathrm{g~cm^{-3}}$.

From our JWST occultation measurements of the C2R ring’s $1^{st}$ and $2^{nd}$ contact radii ($400.3\pm0.1$ and $400.7\pm0.4$ km), we obtain the weighted mean:
\begin{equation}
a = 400.324 \pm 0.097\, \mathrm{km}.
\end{equation}

%To account for the uncertainty in radial width, we uniformly sampled $\delta a$ over the KDE-inferred interval [0.40, 3.25]\, km. 
To account for the uncertainty in the radial width, we uniformly sampled $\delta a$ over the interval defined by the minimum and maximum C2R widths measured across all occultations, [0.10, 3.72] km (see \ref{tab:c2r_equivalent_widths}). A Monte Carlo simulation with $10^5$ realisations yields a median spreading timescale of $1.07 \times 10^7$ s (0.34 yr), with a 1$\sigma$ confidence interval ranging from $6.56 \times 10^6$ s (0.21 yr) to $3.03 \times 10^7$ s (0.96 yr).

%A Monte Carlo simulation with $10^5$ realisations yielded a median spreading timescale of $1.133 \times 10^7$\, s (0.36\, yr), with a 1$\sigma$ confidence interval between $7.392 \times 10^6$\, s (0.23\, yr) and $2.417 \times 10^7$\, s (0.77\, yr).

%To capture the uncertainty in radial width, we sample $\delta a$ uniformly over the KDE-inferred interval [0.40, 3.25]\, km, assuming a fixed bulk density of $\rho=0.79\, \mathrm{g\, cm^{-3}}$ (the mean value of the range reported by Morgado et al. 2021 \cite{Morgado2021}). Monte Carlo sampling with $10^5$ realisations yields a distribution of $T_S$ with a median value of $1.133\times10^7$\, s (0.36\, yr) and a 1$\sigma$ confidence interval between $7.392\times10^6$ s (0.23\, yr) and $2.417\times10^7$\, s (0.77\, yr).}

\subsection{Radiative Transfer Modeling}
\label{subsec:radiative_transfer_modeling}

To explore whether a potential wavelength-dependent effect is responsible for the observed changes in the measured opacities for C1R, we conducted a radiative transfer analysis, assuming that the ring did not change during the observation period (2013-2022) and that all observational chords sampled the same amount of material, after correcting for the opening angle change due to Chariklo's orbital motion around the Sun.
We used a slab-like, two-component model, where we mixed materials to match their common optical properties with the measured occultation normal opacities ($p_N$) or, equivalently, their normal optical depths ($\tau_N$). In our model, the optical depth is obtained as the sum of the optical depths of two materials (two different grain sizes and/or compositions): 
\begin{equation}
    \tau(\lambda) = \tau_1(\lambda)+\tau_2(\lambda)
    \label{eq:tau12}
\end{equation}
which is calculated from the mass densities $\rho_1$ and $\rho_2$, and using the extinction coefficient $\kappa$\,=\,$\kappa^{abs}+\kappa^{scat}$, the sum of the absorption and scattering coefficients:
\begin{equation}
    \tau(\lambda) = [\rho_1\kappa_1(\lambda) + \rho_2\kappa_2(\lambda)]\Delta s
    \label{eq:tau13}
\end{equation}
where $\Delta s$ is the slab's length (the ring's thickness).
A single-component model is obtained if $\rho_2=0$ is chosen.
To calculate the extinction coefficients, we used the OPTOOL package \citep{optool}, which provides direct absorption and scattering coefficients ($\kappa_{abs}$ and $\kappa_{scat}$ in units of cm$^2$/g) for different types of materials and grain sizes. Previous studies based on reflectance spectra between 1998 and 2013 proposed an icy ring system around Chariklo and found that it may contain 20\% water ice and 80\% other materials, such as silicates, organics and small quantities of carbon \citep{Duffard2014}, which motivated our choice of materials to investigate: water ice, carbon and silicates. %%% There is no list of materials before, so we need to insert here 
OPTOOL offers both amorphous and crystalline options for most materials, and we found no significant differences between them in our measurements. Therefore, we used the amorphous option in our analysis. We tested different types of silicates, including pyroxene with varying magnesium content, but found minimal impact on the solutions. Therefore, we used a pyroxene model with 100\% magnesium content, referred to as silicates hereafter.

In general, although $E_P$ is more observationally constrained, compositional information is carried by the opacity estimates. Model calculations (see Equation \ref{eq:tau13}) can provide only optical depth values for direct comparison. However, in $E_P$, this information is lost when it is combined with the radial width of the measurements. 
To compare the observed occultation opacities with our models, we consider the spectral energy distribution $F_*(\lambda)$ of the occulted star and each filter’s transmission curves $R(\lambda)$. For a specific filter $i$, the ratio of occulted and non-occulted photon counts is given by:
\begin{equation}
    r_i = \frac{\int F_*(\lambda)[1-p'(\lambda)] R(\lambda) \lambda d\lambda}{\int F_*(\lambda) R(\lambda) \lambda d\lambda}
\end{equation}
where $p'(\lambda)$ is the wavelength-specific opacity, obtained from the model optical depths as $p'(\lambda)$\,=\,$1-e^{-\tau(\lambda)}$. 

First, we used single-grain models with sizes ranging from 0.2 to 10 $\mu$m. Sizes larger than $\sim$10\,$\mu$m do not show significant differences in extinction properties at different wavelengths in the visible to near-infrared range. We note that for planetary rings, we expect the grain size range to show a power-law distribution with the exponent around three \citep{Brilliantov2015}. However, given the available accuracy of the occultation measurements, our primary goal was to identify the dominant grain sizes. In each case, we determined the best-fit models using least squares minimisation. The $\chi^2$ values were calculated as:
\begin{equation}
    \chi^2 = \sum_i \frac{(r_i-f_i)^2}{df_i^2}
\end{equation}
where $f_i$\,=\,$(1-p_{N,i})$ denotes the normal fractional transmission, with $p_{N,i}$ the measured normal opacity, and $df_i$ its associated uncertainty in the different filters.
%where $f_i$\,=\,$1-p_{N,i}$ \textcolor{red}{is the normal fractional transmission} with $p_{N,i}$ being the normal measured opacity, and $df_i$ is its uncertainty in the different filters. %%% Because we have both rings now to model, this was at wrong place, I replaced it 
After identifying the best-fit grain sizes, we mixed the single-grain solutions using two approaches: (1) mixing the same material with two-grain sizes close to each other and (2) mixing two materials with their preferred grain sizes. The cases with the lowest $\chi^2$ values are considered as our compositional solutions for the rings. 

Here, we perform a thorough analysis of the compatibility of dusty ring models for C1R, assuming a homogeneous layer for each occultation event, composed of a mixture of two materials with different compositions and grain sizes. The pre-JWST data, including the 2017 VLT/Ks measurement at $\sim2.1\,\mu\mathrm{m}$, can be well fitted with a mixture of $\sim10\%$ silicates (olivine or pyroxene) with grain sizes $\lesssim1\,\mu\mathrm{m}$ and $\sim90\%$ water ice with grain sizes between 1--10\,$\mu\mathrm{m}$ (blue curve in Figure ~\ref{fig:c1r_rad_trans}). This fit implies nearly constant opacity across the near-infrared, with increased opacity at shorter visible wavelengths. Comparable fits can also be achieved using mixtures of silicate grains with two distinct size populations, similar to one above. Thus, these data do not strongly constrain the material composition of C1R. However, grains larger than $\sim10\,\mu\mathrm{m}$ would produce nearly wavelength-independent opacity over the full range considered, regardless of composition, suggesting that the grains responsible for most of the extinction in C1R are likely smaller than this $10\,\mu\mathrm{m}$ threshold. 
When JWST data are included, however, no combination of two materials and grain sizes provides a satisfactory fit across all datasets—particularly the JWST F150W2, F322W2, and VLT/Ks bands. The best-fit models in this case (red curve in Figure ~\ref{fig:c1r_rad_trans}) require predominantly silicates with 1--10\,$\mu\mathrm{m}$ grain sizes, but the overall fit quality remains poor. This strongly suggests that the observed opacity variations cannot be explained solely by wavelength dependence. For comparison, spectroscopy of the unresolved Chariklo system indicates a composition of $\leq$20\% water ice mixed with darker materials, probably silicates \citep{Duffard2014}.

\newpage
\section*{Extended Data Figures}
\label{sec:extendeddatafigures}

%\begin{figure}
%    \centering
%    \includegraphics[width=\columnwidth]{figures/chi2_water_silicate_models_v2.pdf}
%    \refstepcounter{extendeddatafig}
%    \label{fig:c1r_c2r_chi2}
%    \caption*{\textbf{\theextendeddatafig:} Model comparison for C2R ring composition. $\chi^2$ values from radiative transfer model fits to the observed C2R opacity across multiple wavelengths. The \textit{nominal particle size distribution} adopted corresponds to pyroxene-rich grains with sizes between $0.2$ and $0.5~\mu$m, which reproduce both the visible and near-infrared opacities. The $\chi^2$ analysis explores mixtures with varying water ice fractions, showing that up to $\sim20\%$ water ice remains consistent with the non-detection in the F322W2 filter. Pure silicate models provide the best fit to the data, but low-ice mixtures are not excluded (see \ref{fig:VLT_vs_JWST} and Figure~\ref{fig:c1rc2r}).}
%\end{figure}

\newpage

\newpage

\newpage
\begin{figure}[ht]
\centering
\includegraphics[width=\columnwidth]{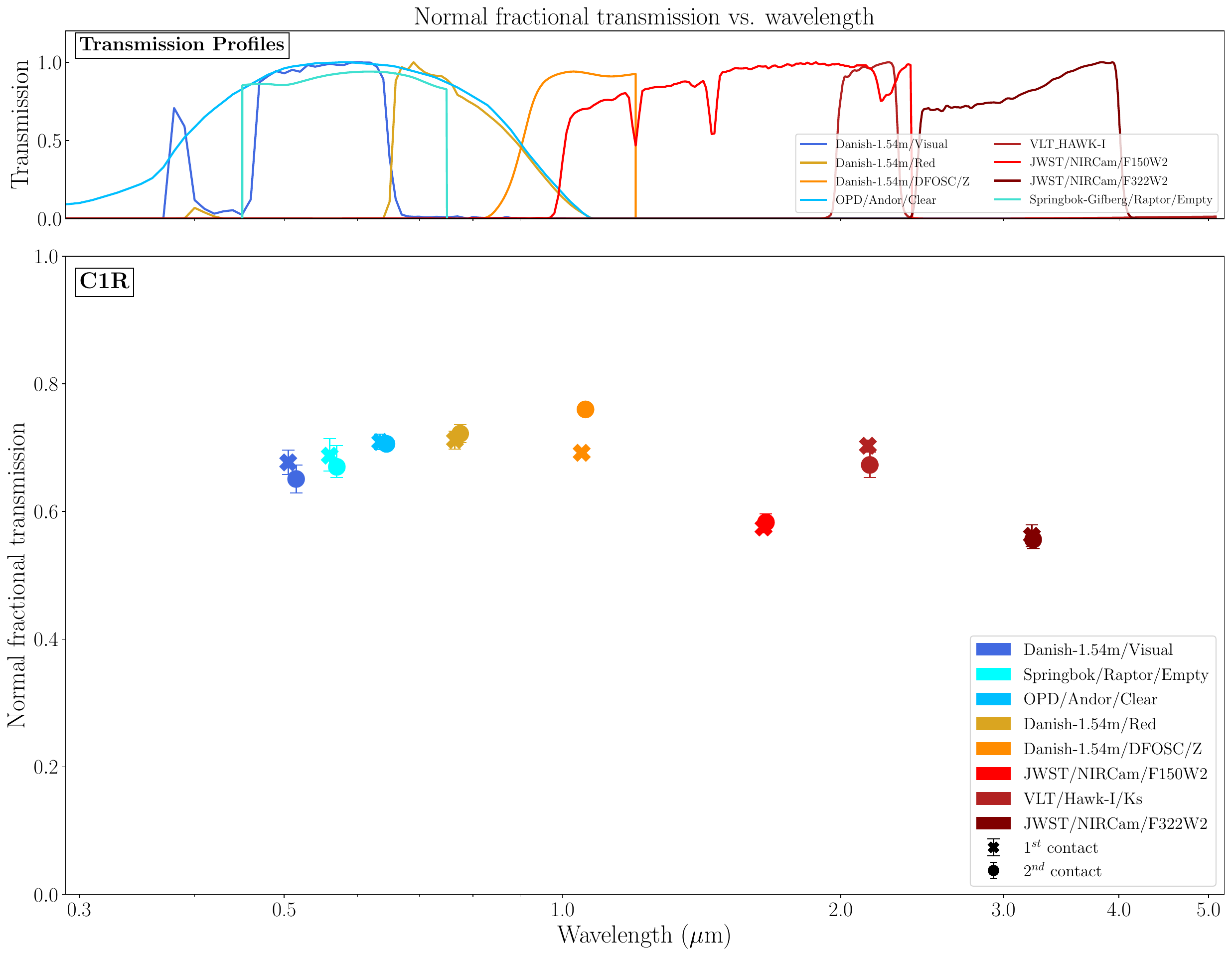}
\refstepcounter{extendeddatafig}
\caption*{\textbf{\theextendeddatafig:}
Wavelength dependence of C1R’s normal fractional transmission $(1 - p_N)$ at the 1$^{\text{st}}$ and 2$^{\text{nd}}$ contact points across multiple stellar occultations. Measurements at the 1$^{\text{st}}$ (“X” symbols) and 2$^{\text{nd}}$ (circle symbols) contacts, obtained from occultations observed between 2013 and 2022, show consistency between ingress and egress within the uncertainties, supporting the use of average $p_N$ values in the analysis. Only events in which C2R was also detected are shown, and all C1R detections included are unambiguous. Detailed observational values are provided in \ref{tab:meas}.
%Wavelength dependence of C1R’s normal fractional transmission ($1 - p_N$) at 1$^{st}$ and 2$^{nd}$ contact points across multiple stellar occultations, based on measurements at the 1$^{\text{st}}$ ('X' symbols) and 2$^{\text{nd}}$ (circle symbols) contact points for occultations observed between 2013 and 2022. The data reveal consistency between ingress and egress values within the uncertainties, justifying the use of their average $p_N$ values in the analysis. Only occultations where C2R was also detected are included. All C1R events shown correspond to unambiguous detections. Detailed observational data are provided in \ref{tab:meas}.
\label{fig:relative_intensity_C1R}}
\end{figure}

\newpage

\begin{figure}[ht!]
\centering

% Panel (a)
\begin{subfigure}[t]{0.80\textwidth}
    \centering
    \includegraphics[width=\linewidth]{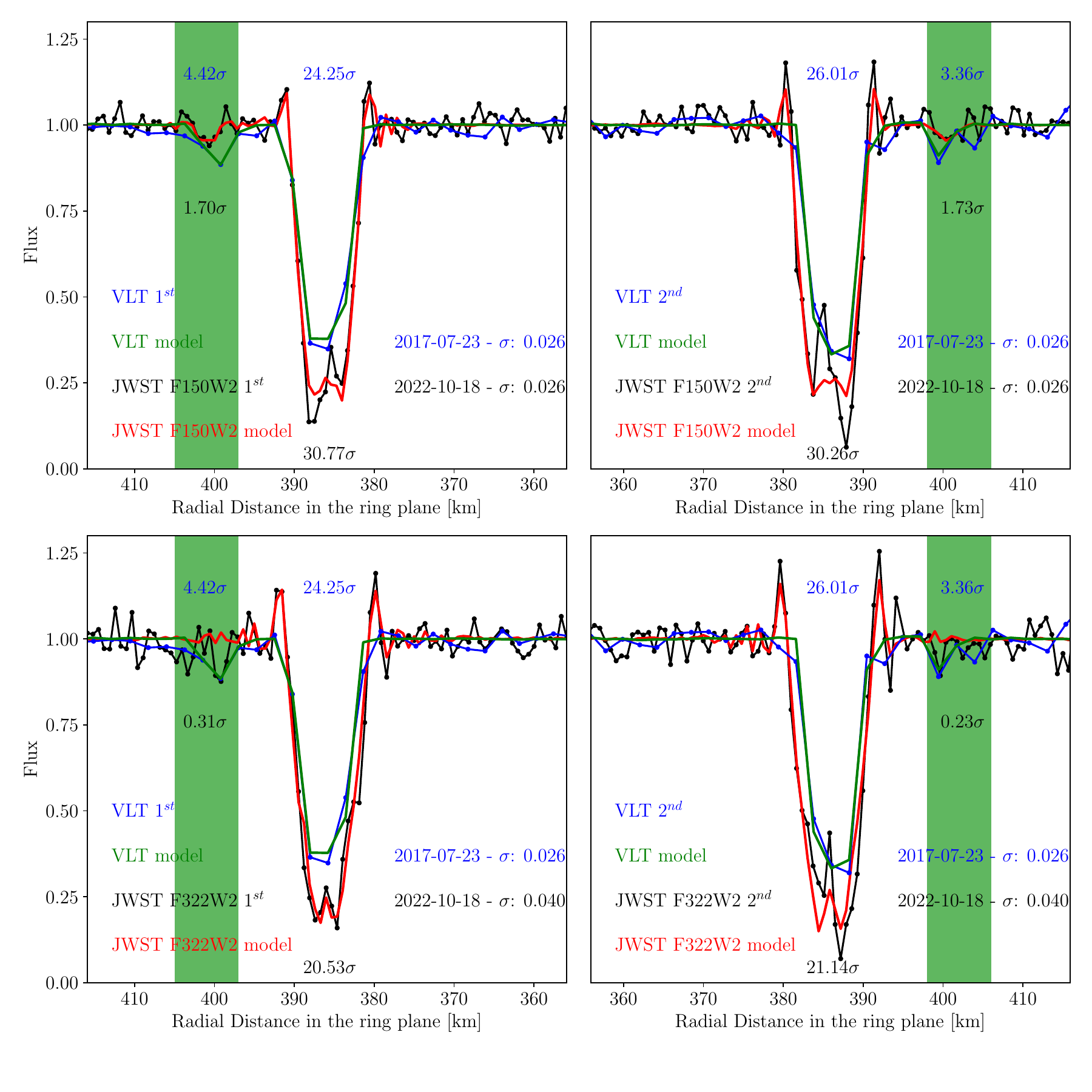}
    \caption*{(a)}
\end{subfigure}

\vspace{0.5cm}

% Panel (b) - Ana kapsayıcı subfigure
\begin{subfigure}[t]{\textwidth}
    \centering
    % Alt görseller
    \begin{subfigure}[t]{0.40\textwidth}
        \centering
        \includegraphics[width=\linewidth]{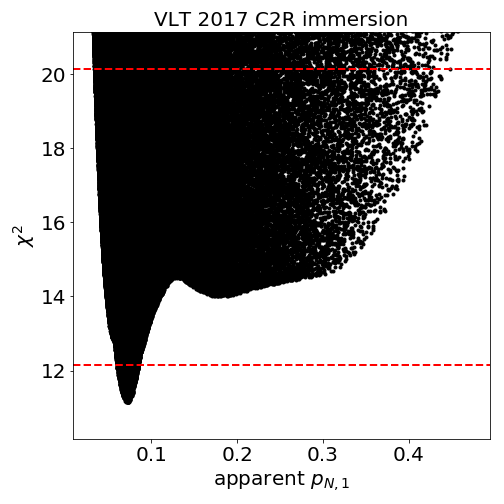}
        \caption*{1$^{\text{st}}$ contact}
    \end{subfigure}
    \hfill
    \begin{subfigure}[t]{0.40\textwidth}
        \centering
        \includegraphics[width=\linewidth]{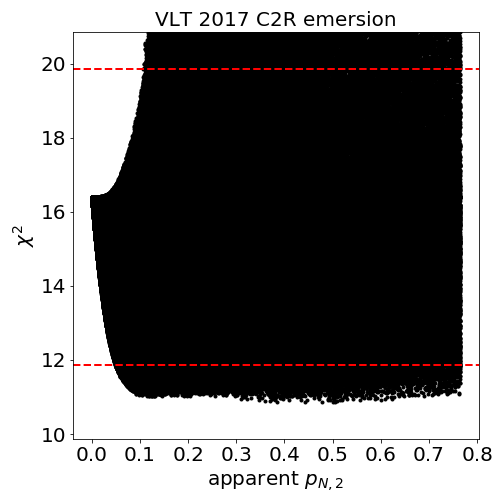}
        \caption*{2$^{\text{nd}}$ contact}
    \end{subfigure}
    \caption*{(b)} % Panel (b) etiketi
\end{subfigure}

\refstepcounter{extendeddatafig}
\caption*{\textbf{\theextendeddatafig:} 
Comparison of JWST (18 Oct. 2022) and VLT HAWK-I (23 Jul. 2017) occultations and VLT-derived constraints on C2R opacity. 
(a) Radial ring profiles of Chariklo’s rings derived from the VLT and JWST  observations. The upper row shows the F150W2 (shorter wavelength) channel of JWST, and the lower row shows the F322W2 (longer wavelength) channel, while the VLT data is identical in both the upper and lower rows. Despite similar SNRs, the VLT profile exhibits larger uncertainties due to the lower spatial resolution of ground-based data. 
(b) Chi-squared distributions from the Monte Carlo fitting of the VLT HAWK-I occultation data in 2017, shown as a function of normal opacity ($p_N$) for the first (left) and second (right) contacts of the C2R ring. Red lines indicate the 1$\sigma$ and 3$\sigma$ confidence levels. The left panel reveals a well-constrained opacity estimate at the first contact, while the right panel illustrates a broader range of consistent $p_N$ values for the second contact, indicating a poorly constrained solution in which the lowest-opacity fit is favoured numerically but not statistically. These distributions help contextualise the differences between the JWST and VLT measurements, highlighting both the evolution of the ring and the limitations inherent to ground-based data.

%Chi-squared distributions from the Monte Carlo fitting of the VLT HAWK-I occultation data in 2017, shown as a function of normal opacity ($p_N$) for the first (left) and second (right) contacts of the C2R ring. Red lines indicate the 1$\sigma$ and 3$\sigma$ confidence levels. The left panel reveals a well-constrained opacity estimate at the first contact. \textcolor{red}{In contrast,} the right panel illustrates a broader range of consistent $p_N$ values for the second contact, indicating a poorly constrained solution where the lowest-opacity fit is favoured numerically but not statistically. %These distributions highlight that the \textcolor{red}{considerable} uncertainty in $p_N$ seen in Figure~\ref{fig:c1rc2r} does not stem from poor data quality but from the inherent degeneracy between ring width and depth in ground-based observations. This degeneracy can only be broken with JWST’s higher spatial resolution, which provides tighter constraints on $p_N$. 
%These plots support the interpretation that the apparent discrepancy between JWST and VLT arises from genuine ring evolution and the limitations of ground-based data, rather than conflicting measurements. For further discussion, see the main text and Methods section.
}
\label{fig:VLT_vs_JWST}
\end{figure}

\newpage
\clearpage
\section*{Extended Data Tables}
\label{sec:extendeddatatables}

\begin{table}[ht!]
    \centering
    \refstepcounter{extendeddatatable}
    \caption*{\textbf{\theextendeddatatable:} Normal opacities ($p_N$) and associated uncertainties for the C1R ring from stellar occultations between 2013 and 2022. Reported values correspond to the 1$^{\text{st}}$ and 2$^{\text{nd}}$ contact points of each event. As both measurements are consistent within their uncertainties, the average $p_N$ was adopted in the analysis. Only events in which the C2R ring could also be detected are included. All entries list the nominal $p_N$ value along with the corresponding lower and upper uncertainties. All C1R detections are considered robust.
    }
    \label{tab:meas}
    \begin{tabular}{llrccc}
    \hline
    Telescope/Instrument/Filter & Date & Ring & $p_{N,1} \pm dp_{N,1}$ & $p_{N,2} \pm dp_{N,2}$ & $p_{N} \pm dp_{N}$ \\
    \hline
    Danish-1.54m/DFOSC/Z$^{a}$ & 03/06/2013 & C1R & $0.308 \pm 0.003$ & $0.240 \pm 0.004$ & $0.284 \pm 0.002$ \\
    Springbok/Raptor/Empty$^{a}$  & 29/04/2014 & C1R & $0.3125^{+0.024}_{-0.027}$ & $0.33^{+0.017}_{-0.033}$ & $0.321^{+0.021}_{-0.030}$ \\
    Danish-1.54m/Red$^{b}$ & 23/07/2017 & C1R & $0.288 \pm 0.014$ & $0.278 \pm 0.014$ & $0.283 \pm 0.015$ \\
    Danish-1.54m/Visual$^{b}$ & 23/07/2017 & C1R & $0.323 \pm 0.019$ & $0.349 \pm 0.022$ & $0.336 \pm 0.026$ \\
    OPD/Andor/Clear$^{b}$ & 23/07/2017 & C1R & $0.291 \pm 0.012$ & $0.294 \pm 0.009$ & $0.293 \pm 0.012$ \\
    VLT/Hawk-I/K$s$$^{b}$ & 23/07/2017 & C1R & $0.297 \pm 0.009$ & $0.327 \pm 0.020$ & $0.312 \pm 0.027$ \\
    JWST/NIRCam/F150W2$^{c}$ & 18/10/2022 & C1R & $0.425 \pm 0.008$ & $0.417 \pm 0.013$ & $0.421 \pm 0.016$ \\
    JWST/NIRCam/F322W2$^{c}$ & 18/10/2022 & C1R & $0.438 \pm 0.017$ & $0.444 \pm 0.014$ & $0.441 \pm 0.022$ \\
    \hline
    \end{tabular}
    \textbf{References.}: \textit{a.} \cite{Berard2017}, \textit{b.} \cite{Morgado2021}, \textit{c.} This work. \\
\end{table}

\begin{table}[ht!]
    \centering
    \refstepcounter{extendeddatatable}
    \caption*{\textbf{\theextendeddatatable:} Radial widths ($W_r$) and equivalent widths ($E_P$) for the C1R ring derived from stellar occultations. Reported values correspond to the 1$^{\text{st}}$ and 2$^{\text{nd}}$ contact points of each event, with asymmetric error bars (see Figure~\ref{fig:ep_over_time_c1r_c2r}, Panel (a)).}
    \begin{tabular}{l l c c c c}
    \hline
    Telescope/Instrument/Filter & Date &
    \begin{tabular}{c}
    $W_{r,1}$ \\
    (km)
    \end{tabular} &
    \begin{tabular}{c}
    $W_{r,2}$ \\
    (km)
    \end{tabular} &
    \begin{tabular}{c}
    $E_{p,1}$ \\
    (km)
    \end{tabular} &
    \begin{tabular}{c}
    $E_{p,2}$ \\
    (km)
    \end{tabular} \\
    \hline

Danish-1.54m/DFOSC/Z$^{a}$ & 2013-06-03 &
$6.160^{+0.110}_{-0.110}$ & $7.140^{+0.040}_{-0.040}$ &
$1.9^{+0.022}_{-0.022}$ & $1.73^{+0.023}_{-0.023}$ \\

VLT/Hawk-I/Ks$^{a}$ & 2014-02-16 &
$5.316^{+0.868}_{-1.916}$ & $4.833^{+1.667}_{-0.476}$ &
$1.996^{+0.092}_{-0.031}$ & $2.04^{+0.36}_{-0.14}$ \\

SAAO/SHOC$^{a}$ & 2014-04-29 &
$5.680^{+0.200}_{-0.200}$ & $6.625^{+0.200}_{-0.200}$ &
$1.88^{+0.22}_{-0.12}$ & $1.695^{+0.175}_{-0.115}$ \\

Springbok/Raptor/Empty$^{a}$ & 2014-04-29 &
$5.575^{+0.398}_{-0.398}$ & $6.750^{+0.480}_{-0.210}$ &
$1.8^{+0.122}_{-0.143}$ & $2.595^{+0.148}_{-0.166}$ \\

Hakos/Raptor/Clear$^{b}$ & 2017-04-09 &
-- & $5.642^{+0.205}_{-0.205}$ &
-- & $2.319^{+0.23}_{-0.23}$ \\

WabiLodge/Raptor/Clear$^{b}$ & 2017-04-09 &
$7.056^{+0.258}_{-0.258}$ & $5.725^{+0.167}_{-0.167}$ &
$1.7^{+0.318}_{-0.318}$ & $1.866^{+0.314}_{-0.314}$ \\

WeaverRL/Raptor/Clear$^{b}$ & 2017-04-09 &
$7.591^{+0.717}_{-0.717}$ & $5.712^{+0.581}_{-0.581}$ &
$2.179^{+0.364}_{-0.364}$ & $2.319^{+0.381}_{-0.381}$ \\

Tivoli/Raptor/Clear$^{b}$ & 2017-06-22 &
$3.872^{+1.269}_{-2.034}$ & $3.909^{+1.008}_{-1.561}$ &
$1.24^{+0.388}_{-0.334}$ & $1.571^{+0.489}_{-0.382}$ \\

WdhBackes/Raptor/Clear$^{b}$ & 2017-06-22 &
$7.459^{+2.267}_{-3.973}$ & $9.987^{+2.967}_{-2.690}$ &
$2.045^{+1.514}_{-0.381}$ & $1.746^{+0.56}_{-0.416}$ \\

Onduruquea/Raptor/Clear$^{b}$ & 2017-06-22 &
$5.891^{+0.441}_{-0.636}$ & $5.550^{+0.532}_{-0.486}$ &
$2.121^{+0.179}_{-0.351}$ & $1.744^{+0.239}_{-0.26}$ \\

Outeniqua/Raptor/Clear$^{b}$ & 2017-06-22 &
$8.266^{+1.972}_{-2.750}$ & $8.104^{+1.706}_{-1.880}$ &
$3.391^{+2.654}_{-1.081}$ & $3.989^{+2.919}_{-1.37}$ \\

WdhMeza/ASI178MM/Clear$^{b}$ & 2017-06-22 &
$8.189^{+3.029}_{-6.027}$ & $3.731^{+2.011}_{-0.731}$ &
$1.577^{+1.34}_{-0.59}$ & $2.816^{+0.81}_{-1.172}$ \\

Hakos/Raptor/Clear$^{b}$ & 2017-06-22 &
$5.095^{+0.273}_{-0.197}$ & $5.558^{+0.365}_{-0.426}$ &
$1.74^{+0.202}_{-0.176}$ & $1.739^{+0.125}_{-0.25}$ \\

Danish-1.54m/Red$^{b}$ & 2017-07-23 &
$5.779^{+0.162}_{-0.162}$ & $8.326^{+0.246}_{-0.246}$ &
$1.664^{+0.13}_{-0.13}$ & $2.315^{+0.188}_{-0.188}$ \\

VLT/Hawk-I/Ks$^{b}$ & 2017-07-23 &
$6.596^{+0.262}_{-0.262}$ & $6.524^{+0.36}_{-0.36}$ &
$1.959^{+0.14}_{-0.14}$ & $2.133^{+0.255}_{-0.255}$ \\

SSO/Andor/g/I+z$^{b}$ & 2017-07-23 &
$7.920^{+14.125}_{-3.701}$ & $4.760^{+12.671}_{-0.122}$ &
$3.118^{+0.861}_{-1.287}$ & $3.565^{+0.09}_{-1.658}$ \\

ElRodeo/MerlinKite/Clear$^{b}$ & 2017-07-23 &
$7.920^{+14.125}_{-3.701}$ & $15.103^{+4.076}_{-3.993}$ &
$3.118^{+0.861}_{-1.287}$ & $3.004^{+1.295}_{-1.346}$ \\

ElSalvador/Raptor/Clear$^{b}$ & 2017-07-23 &
$7.811^{+1.367}_{-0.824}$ & $5.390^{+7.818}_{-1.695}$ &
$3.033^{+1.428}_{-0.57}$ & $1.307^{+0.854}_{-0.53}$ \\

Danish-1.54m/Visual$^{b}$ & 2017-07-23 &
$6.064^{+0.179}_{-0.179}$ & $8.191^{+0.378}_{-0.378}$ &
$1.959^{+0.176}_{-0.176}$ & $2.859^{+0.32}_{-0.32}$ \\

LaSilla1m/Raptor/Clear$^{b}$ & 2017-07-23 &
$6.035^{+0.557}_{-0.557}$ & $9.107^{+0.398}_{-0.398}$ &
$1.811^{+0.352}_{-0.352}$ & $2.677^{+0.317}_{-0.317}$ \\

OPD/Andor/Clear$^{b}$ & 2017-07-23 &
$6.456^{+0.134}_{-0.134}$ & $8.920^{+0.141}_{-0.141}$ &
$1.879^{+0.118}_{-0.118}$ & $2.622^{+0.123}_{-0.123}$ \\

TolarGrande/Raptor/Clear$^{b}$ & 2017-07-23 &
$8.670^{+1.635}_{-1.635}$ & $6.068^{+1.229}_{-1.229}$ &
$1.795^{+0.782}_{-0.782}$ & $2.057^{+0.92}_{-0.92}$ \\

IncaDeOro/Raptor/Clear$^{b}$ & 2017-07-23 &
$6.614^{+2.553}_{-4.898}$ & $3.915^{+1.020}_{-1.561}$ &
$1.51^{+0.621}_{-0.714}$ & $1.542^{+1.077}_{-0.423}$ \\

OPD/Andor/Clear$^{b}$ & 2017-08-24 &
$9.807^{+1.724}_{-5.492}$ & $6.465^{+10.634}_{-2.946}$ &
$2.103^{+1.769}_{-0.296}$ & $3.814^{+1.965}_{-2.148}$ \\

Glenlee/Watec/Clear$^{b}$ & 2020-06-19 &
$40.396^{+8.286}_{-8.286}$ & $6.465^{+10.634}_{-2.946}$ &
$7.192^{+2.381}_{-1.692}$ & $3.814^{+1.956}_{-2.148}$ \\

JWST/NIRCam/F150W2$^{c}$ & 2022-10-18 &
$7.040^{+0.070}_{-0.070}$ & $7.600^{+0.100}_{-0.100}$ &
$2.99^{+0.05}_{-0.05}$ & $3.17^{+0.09}_{-0.09}$ \\

JWST/NIRCam/F322W2$^{c}$ & 2022-10-18 &
$7.020^{+0.200}_{-0.200}$ & $7.500^{+0.100}_{-0.100}$ &
$3.08^{+0.09}_{-0.09}$ & $3.33^{+0.1}_{-0.1}$ \\
    \hline
    \end{tabular}
    \label{tab:c1r_equivalent_widths}
    \textbf{References.}: \textit{a.} \cite{Berard2017}, \textit{b.} \cite{Morgado2021}, \textit{c.} This work. \\
\end{table}

\begin{table}[ht!]
    \centering
    \refstepcounter{extendeddatatable}
    \caption*{\textbf{\theextendeddatatable:} Radial widths ($W_r$) and equivalent widths ($E_P$) for the C2R ring derived from stellar occultations. Reported values correspond to the 1$^{\text{st}}$ and 2$^{\text{nd}}$ contact points of each event, with asymmetric uncertainties (see Figure~\ref{fig:ep_over_time_c1r_c2r}, Panel (b)).}
    
    \begin{tabular}{l l c c c c}
    \hline
    Telescope/Instrument/Filter & Date &
    \begin{tabular}{c}
    $W_{r,1}$ \\
    (km)
    \end{tabular} &
    \begin{tabular}{c}
    $W_{r,2}$ \\
    (km)
    \end{tabular} &
    \begin{tabular}{c}
    $E_{p,1}$ \\
    (km)
    \end{tabular} &
    \begin{tabular}{c}
    $E_{p,2}$ \\
    (km)
    \end{tabular} \\
    \hline

Danish-1.54m/DFOSC/Z$^{a}$ & 03/06/2013 &
$3.380^{+1.424}_{-1.797}$ & $3.231^{+0.899}_{-1.124}$ &
$0.168^{+0.02}_{-0.02}$ & $0.228^{+0.02}_{-0.02}$ \\

Springbok/Raptor/Empty$^{a}$ & 29/04/2014 &
$0.34^{+1.37}_{-0.24}$ & $0.6^{+1.7}_{-0.1}$ &
$0.125^{+0.076}_{-0.064}$ & $0.253^{+0.079}_{-0.069}$ \\

Gifberg/Raptor/Empty$^{a}$ & 29/04/2014 &
$0.522^{+0.227}_{-0.399}$ & $0.181^{+0.008}_{-0.091}$ &
$0.090^{+0.039}_{-0.000}$ & $0.129^{+0.000}_{-0.039}$ \\

Danish-1.54m/Visual$^{b}$ & 23/07/2017 &
$0.277^{+0.640}_{-0.129}$ & $0.258^{+0.161}_{-0.072}$ &
$0.123^{+0.036}_{-0.026}$ & $0.154^{+0.039}_{-0.034}$ \\

Danish-1.54m/Red$^{b}$ & 23/07/2017 &
$0.191^{+1.233}_{-0.073}$ & $0.801^{+0.440}_{-0.559}$ &
$0.095^{+0.021}_{-0.016}$ & $0.127^{+0.029}_{-0.035}$ \\

OPD/Andor/Clear$^{b}$ & 23/07/2017 &
$0.631^{+0.545}_{-0.185}$ & $0.095^{+0.015}_{-0.010}$ &
$0.096^{+0.015}_{-0.012}$ & $0.072^{+0.010}_{-0.009}$ \\

VLT/Hawk-I/K$s^{b}$ & 23/07/2017 &
$3.716^{+0.631}_{-0.527}$ & $0.159^{+3.110}_{-0.097}$ &
$0.168^{+0.016}_{-0.020}$ & $0.105^{+0.096}_{-0.061}$ \\

JWST/NIRCam/F150W2$^{c}$ & 18/10/2022 &
$1.009^{+0.483}_{-0.623}$ & $2.273^{+2.624}_{-1.411}$ &
$0.049^{+0.016}_{-0.010}$ & $0.039^{+0.038}_{-0.025}$ \\
    \hline
    \end{tabular}

    \label{tab:c2r_equivalent_widths}
    \textbf{References.}: \textit{a.} \cite{Berard2017}, \textit{b.} \cite{Morgado2021}, \textit{c.} This work.
\end{table}

\newpage
\clearpage
\bibliography{sn-bibliography_nat-comm_v4}

\section*{Acknowledgements}

This work is based on observations made with the NASA/ESA/CSA James Webb Space Telescope. The data were obtained from the Mikulski Archive for Space Telescopes at the Space Telescope Science Institute, which is operated by the Association of Universities for Research in Astronomy, Inc., under NASA contract NAS 5-03127 for JWST. These observations are associated with program \#1271. All the JWST data used in this paper can be found in MAST: \url{https://doi.org/10.17909/qaaf-8q45}. This work has made use of data from the European Space Agency (ESA) mission {\it Gaia} (\url{https://www.cosmos.esa.int/gaia}), processed by the {\it Gaia} Data Processing and Analysis Consortium (DPAC, \url{https://www.cosmos.esa.int/web/gaia/dpac/consortium}). Funding for the DPAC has been provided by national institutions, in particular, the institutions participating in the {\it Gaia} Multilateral Agreement. P.S-S. acknowledges financial support from the Spanish I+D+i project PID2022-139555NB-I00 (TNO-JWST) funded by MCIN/AEI/10.13039/501100011033. Y.K. is forever thankful to TÜBİTAK National Observatory (1997-2024) for its unwavering support of his career and science. The authors acknowledge financial support from the Severo Ochoa grant CEX2021-001131-S funded by MICIU/AEI/ 10.13039/501100011033. M.A. thanks CNPq grants 427700/2018-3, 310683/2017-3 and 473002/2013-2. A.R.G.J. thanks the financial support of São Paulo Research Foundation (FAPESP, proc. 2016/24561-0 and 2018/11239-8). C.L.P. is thankful for the support of the Coordenação de Aperfeiçoamento de Pessoal de Nível Superior - Brazil (CAPES) and Fundação de Amparo à Pesquisa do Estado do Rio de Janeiro - FAPERJ/DSC-10 (E26/204.141/2022). HBH and SNM also acknowledge support from NASA JWST Interdisciplinary Scientist grant 21-SMDSS21-0013. CK was supported by the K-138962 and  TKP2021-NKTA-64 grants of the National Research, Development and Innovation Office (NKFIH), Hungary. We sincerely thank both anonymous referees for their exceptional and insightful contributions, which significantly improved the clarity and quality of this manuscript.

\end{document}